\documentclass[twocolumn,twocolappendix]{aastex631}

\newcommand{\rhigh}{R_{\rm high}}

\newcommand{\spin}{a_{\ast}}
\newcommand{\PA}{\mathrm{PA}}

\usepackage{hyperref}

\shorttitle{Rotation in Event Horizon Telescope Movies}
\shortauthors{Conroy et al.}

\graphicspath{{./}{figures/}}

\begin{document}

\title{Rotation in Event Horizon Telescope Movies}

\author[0000-0003-2886-2377]{Nicholas S. Conroy}
\affiliation{Department of Astronomy, University of Illinois at Urbana-Champaign, 1002 West Green Street, Urbana, IL 61801, USA}

\author[0000-0002-5518-2812]{Michi Baub\"ock}
\affiliation{Department of Physics, University of Illinois at Urbana-Champaign, 1110 West Green Street, Urbana, IL 61801, USA}
\affiliation{Illinois Center for the Advanced Study of the Universe, University of Illinois at Urbana-Champaign, 1110 West Green St., Urbana, IL 61801, USA}

\author[0000-0001-6765-877X]{Vedant Dhruv}
\affiliation{Department of Physics, University of Illinois at Urbana-Champaign, 1110 West Green Street, Urbana, IL 61801, USA}
\affiliation{Illinois Center for the Advanced Study of the Universe, University of Illinois at Urbana-Champaign, 1110 West Green St., Urbana, IL 61801, USA}

\author[0000-0002-3350-5588]{Daeyoung Lee}
\affiliation{Department of Physics, University of Illinois at Urbana-Champaign, 1110 West Green Street, Urbana, IL 61801, USA}

\author[0000-0002-3351-760X]{Avery E. Broderick}
\affiliation{Perimeter Institute for Theoretical Physics, 31 Caroline Street North, Waterloo, ON, N2L 2Y5, Canada}
\affiliation{Department of Physics and Astronomy, University of Waterloo, 200 University Avenue West, Waterloo, ON, N2L 3G1, Canada}
\affiliation{Waterloo Centre for Astrophysics, University of Waterloo, Waterloo, ON N2L 3G1 Canada}

\author[0000-0001-6337-6126]{Chi-kwan Chan}
\affiliation{Steward Observatory and Department of Astronomy, University of Arizona, 933 N. Cherry Ave., Tucson, AZ 85721, USA}
\affiliation{Data Science Institute, University of Arizona, 1230 N. Cherry Ave., Tucson, AZ 85721, USA}
\affiliation{Program in Applied Mathematics, University of Arizona, 617 N. Santa Rita, Tucson, AZ 85721, USA}

\author[0000-0002-3586-6424]{Boris Georgiev}
\affiliation{Perimeter Institute for Theoretical Physics, 31 Caroline Street North, Waterloo, ON, N2L 2Y5, Canada}
\affiliation{Department of Physics and Astronomy, University of Waterloo, 200 University Avenue West, Waterloo, ON, N2L 3G1, Canada}
\affiliation{Waterloo Centre for Astrophysics, University of Waterloo, Waterloo, ON N2L 3G1 Canada}

\author[0000-0002-2514-5965]{Abhishek V. Joshi}
\affiliation{Department of Physics, University of Illinois at Urbana-Champaign, 1110 West Green Street, Urbana, IL 61801, USA}
\affiliation{Illinois Center for the Advanced Study of the Universe, University of Illinois at Urbana-Champaign, 1110 West Green St., Urbana, IL 61801, USA}

\author[0000-0002-0393-7734]{Ben Prather}
\affiliation{Department of Physics, University of Illinois at Urbana-Champaign, 1110 West Green Street, Urbana, IL 61801, USA}
\affiliation{Los Alamos National Lab, Los Alamos, NM, 87545}

\author[0000-0001-7451-8935]{Charles F. Gammie}
\affiliation{Department of Physics, University of Illinois at Urbana-Champaign, 1110 West Green Street, Urbana, IL 61801, USA}
\affiliation{Department of Astronomy, University of Illinois at Urbana-Champaign, 1002 West Green Street, Urbana, IL 61801, USA}
\affiliation{NCSA, University of Illinois at Urbana-Champaign, 1205 W. Clark St., Urbana, IL 61801, USA}
\affiliation{Illinois Center for the Advanced Study of the Universe, University of Illinois at Urbana-Champaign, 1110 West Green St., Urbana, IL 61801, USA}
 
\begin{abstract}

The Event Horizon Telescope (EHT) has produced images of M87* and Sagittarius A*, and will soon produce time sequences of images, or movies.  In anticipation of this, we describe a technique to measure the rotation rate, or pattern speed $\Omega_p$, from movies using an autocorrelation technique.  We validate the technique on Gaussian random field models with a known rotation rate and apply it to a library of synthetic images of Sgr A* based on general relativistic magnetohydrodynamics simulations.  We predict that EHT movies will have $\Omega_p \approx 1^\circ$ per $GMc^{-3}$, which is of order $15\%$ of the Keplerian orbital frequency in the emitting region. We can plausibly attribute the slow rotation seen in our models to the pattern speed of inward-propagating spiral shocks. We also find that $\Omega_p$ depends strongly on inclination. Application of this technique will enable us to compare future EHT movies with the clockwise rotation of Sgr A* seen in near-infrared flares by GRAVITY.  Pattern speed analysis of future EHT observations of M87* and Sgr A* may also provide novel constraints on black hole inclination and spin, as well as an independent measurement of black hole mass.

\end{abstract}

\section{Introduction} 
\label{sec:intro}

The Event Horizon Telescope (EHT) has imaged the black hole Sagittarius A* (Sgr A*) at the heart of our own galaxy (\citealt{SgrAPaperI}) and the black hole M87* at the center of M87 (\citealt{M87PaperI}) at event horizon scale resolution. These images were made by combining data from an array of radio telescopes using a technique called very long baseline interferometry (VLBI). For M87*, key science results include a mass measurement that is consistent with estimates based on stellar kinematics (\citealt{Gebhardt_2011}). For Sgr A*, key results include a mass measurement that is consistent with earlier, more-precise measurements based on individual stellar orbits \citep{Schoedel_2002, Ghez_2003, Ghez_2008, Do_2019, GRAVITY_2019, GRAVITY_2020_precession}. 

\begin{figure*}
\gridline{\fig{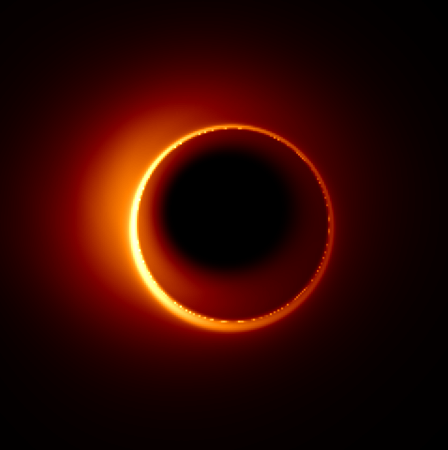}{0.30\textwidth}{}
          \fig{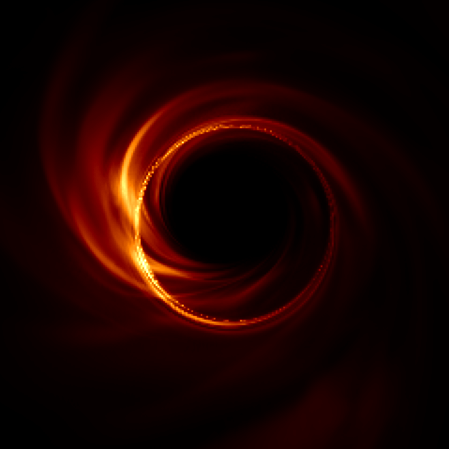}{0.30\textwidth}{}
          \fig{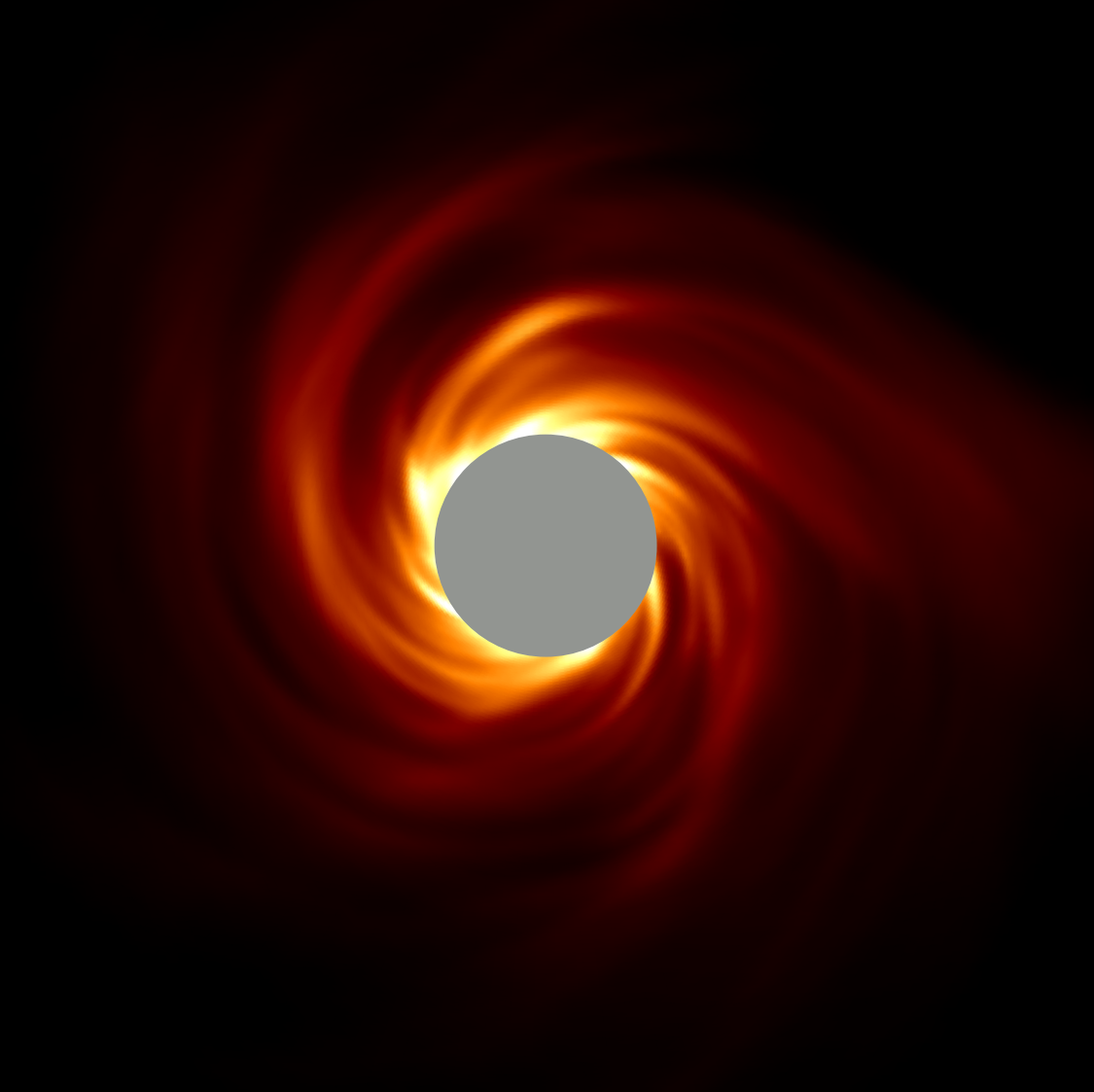}{0.30\textwidth}{}
          }
    \caption{
        Example time-averaged 230 GHz image (left) from a fiducial Sgr A* model (MAD, $\spin = 0.5$, $i = 30^\circ$, $\rhigh = 160$), a snapshot 230 GHz image (center), and a plot of pressure averaged around the black hole's equatorial plane from the same time slice as the snapshot (right). In the right panel the region inside $r = 2\,GMc^{-2}$ is grayed out as it contributes relatively little emission.  Notice that the nonaxisymmetric, time-variable structures form trailing spirals that are visible in both the snapshot and GRMHD pressure field.  The field of view in the left and center panels is $20 GM/(c^2 D)$, and the extent of the right panel is $20\,GMc^{-2}$.
        \label{fig:example}
    }
\end{figure*}

Interpretations of EHT data have relied heavily on time-dependent general relativistic magnetohydrodynamics (GRMHD) models, which are remarkably consistent with the data \citep{M87PaperV, M87PaperVIII, SgrAPaperV, Wong_2022}.  In M87*, GRMHD models predicted \citep{M87PaperV} that the angle between the brightness maximum on the ring and the large-scale jet in M87* observed in 2017, $\sim 150^\circ$, was an outlier, and that an angle closer to $\sim 90^\circ$ would be more frequently observed.  This is consistent with data from other epochs \citep{Wielgus_2020}. 
In Sgr A*, however, GRMHD models predict a source-integrated variability that is a factor of two larger than observed \citep{SgrAPaperV, Wielgus_2022}, focusing interest on the origins of variability in GRMHD models.

Variability is likely to become a focal point for EHT science.  The EHT is developing the ability to revisit sources regularly, enabling movies of M87*, while also expanding its baseline coverage, enabling movies of Sgr A* \citep{ngEHT_whitepaper, ngEHT_whitepaper2}.
What might movies reveal about both of the resolved EHT sources?

The hot spot model is a common starting point for understanding nonaxisymmetric variability.  In the simplest version of this model, a hot spot moves freely on a circular orbit in the equatorial plane of the black hole (e.g. \citealt{Broderick_2006}, \citealt{Emami_2022}, \citealt{Wielgus_2022_hotspots}).  Assuming emission arises near $x \equiv R c^2/(GM) \sim 4$, as it does in GRMHD-based models \citep[see, e.g., Figure 4 of][]{M87PaperV}, then we expect the hot spot to orbit at the circular geodesic, or Keplerian, frequency $\Omega_K = (GM/c^{-3})^{-1} (x^{3/2} + \spin)^{-1}$. For a face-on black hole with spin $\spin \equiv Jc/GM^2 = 0$, $\Omega_K \approx 7 (x/4)^{-3/2}$ degrees per $GMc^{-3}$. 
This frequency is an important point of comparison for variability in EHT movies.  

GRMHD models do not show freely orbiting hot spots.  Instead, they tend to show transient spiral features.  Figure \ref{fig:example} shows a time-averaged image from a GRMHD model in the left panel next to a typical snapshot from the same model in the center panel.  Evidently the nonaxisymmetric, time-dependent emission is concentrated in spiral features.  The underlying plasma is subject to pressure gradient forces and magnetic forces, so the plasma need not move on geodesics.  Strongly magnetized models (called magnetically arrested disks, or ``MADs'') tend to show rotation that is sub-Keplerian, while weakly magnetized models (called standard and normal evolution, or``SANEs'') are closer to Keplerian.  Radial velocities are typically close to the sound speed, particularly in models where the emission peaks inside the innermost stable circular orbit, in the so-called plunging region.  The plasma motion is not well described by circular orbits.   

\begin{figure*}[!t]
    \plottwo{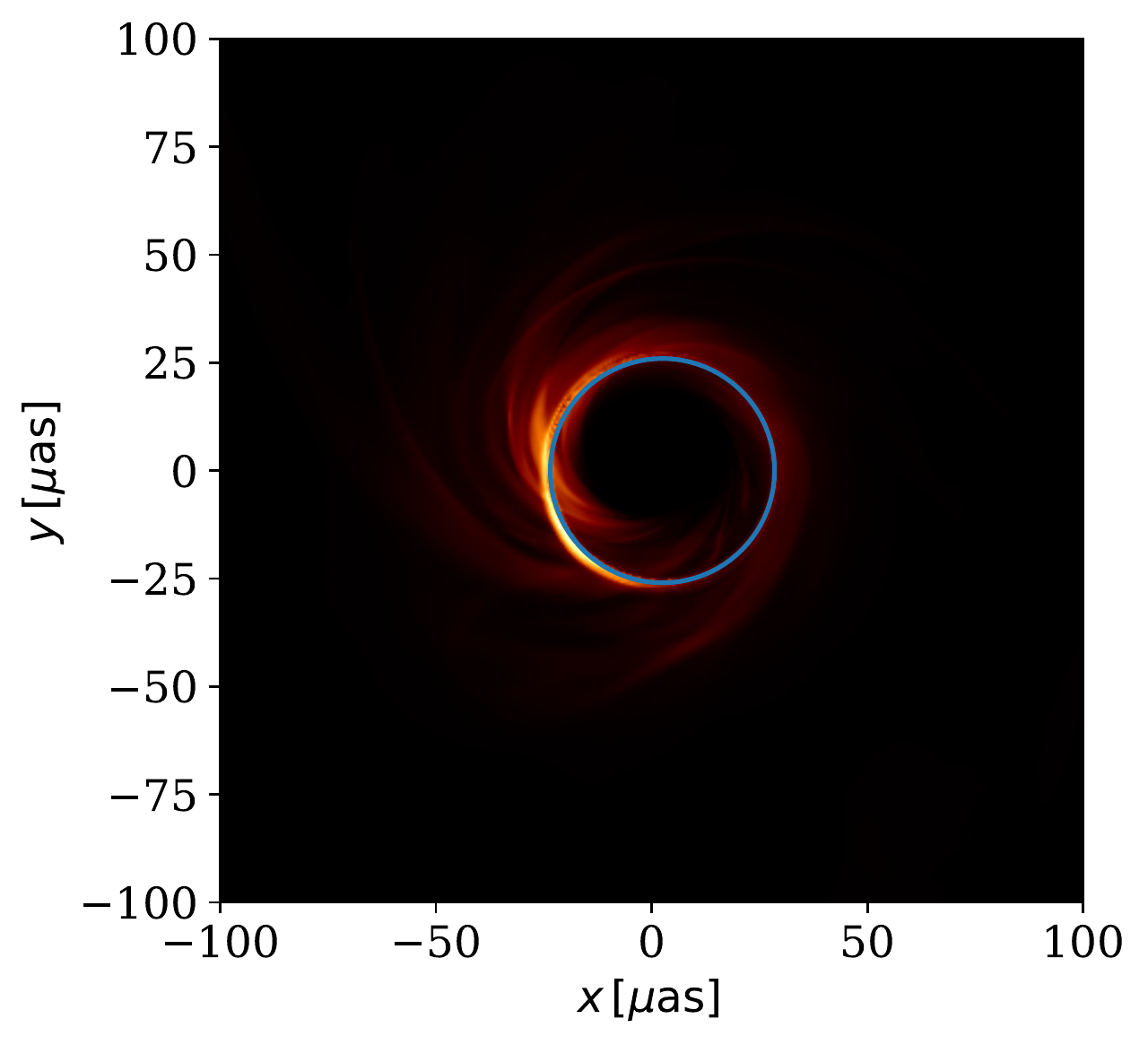}{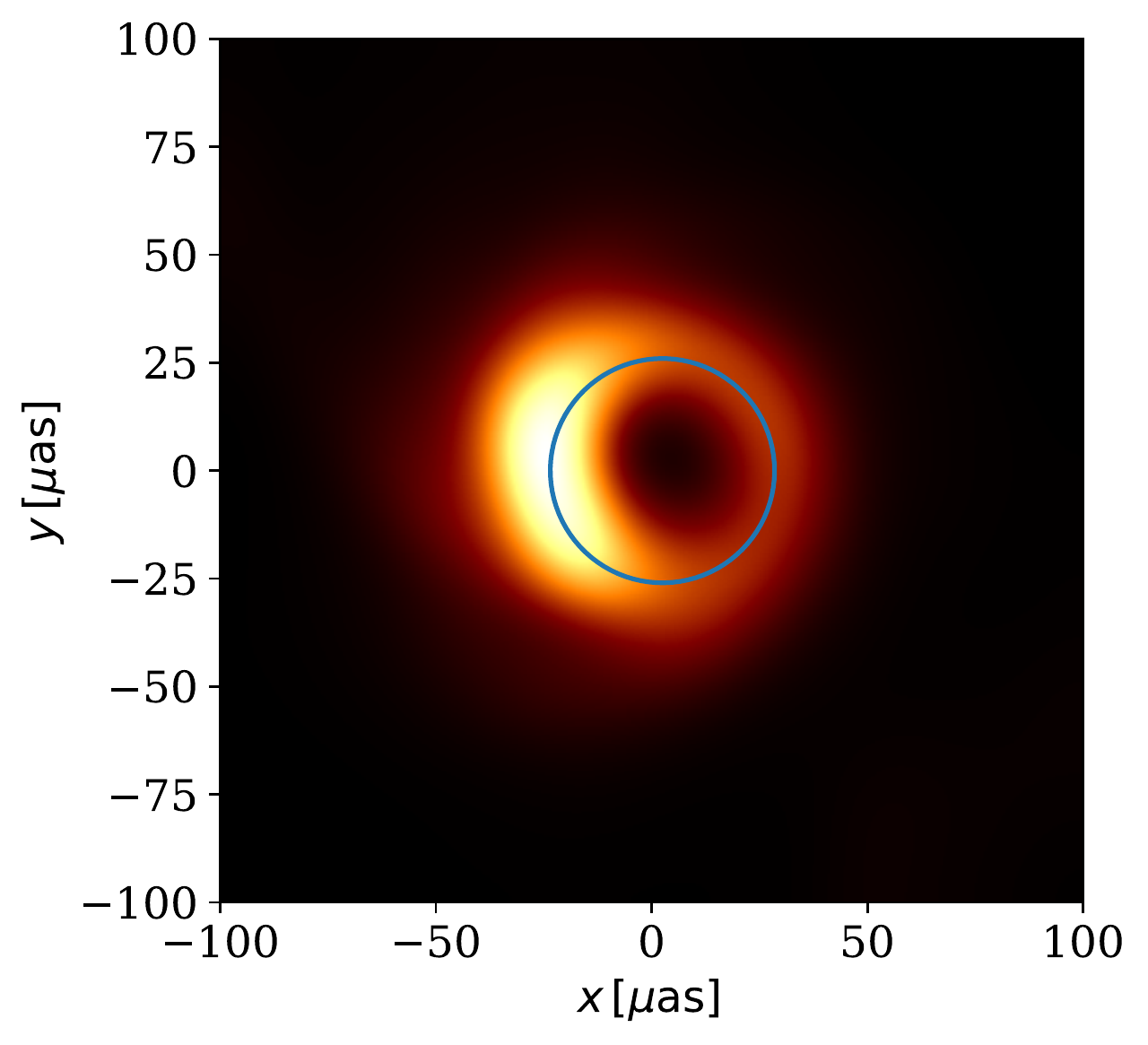}
    \caption{
        An unblurred (left) and blurred (right) synthetic image.  The blue ring shows the sampling circle from Equation~(\ref{eq:ringdef}).  The image is drawn from the fiducial Sgr A* model (MAD, $\spin = 0.5$, $i = 30^{\circ}$, $R_{high} = 160$). 
        \label{fig:ringlocation}
    }
\end{figure*}

The motion of the spiral features seen in Figure \ref{fig:example} may be detectable in EHT movies.  In this paper, we define and evaluate the pattern speed $\Omega_p$, which is a measure of rotation in EHT movies. Our analysis is based on synthetic GRMHD data from the Illinois Sgr A* model library, which was run using KHARMA, an ideal nonradiative GRMHD code\footnote{KHARMA (B. Prather et al. 2023, in preparation) is a GPU-enabled version of HARM \citep{Gammie_03}.  It is publicly available at \url{https://github.com/AFD-Illinois/kharma}} and imaged with {\tt ipole} \citep{Mosc_2018}.  The model library movies have an angular resolution of $0.5\,\mu$as and a time resolution of $5\,GMc^{-3}$ between images.  Each model comprises $3 \times 10^3$ images evenly spaced between time $1.5 \times 10^4$ to time $3 \times 10^4\,GMc^{-3}$ after their initialization with a magnetized torus.  In Sgr A*, where $GMc^{-3} \simeq 20 \, \mathrm{s}$, the time between frames is $100 \, \mathrm{s}$ and the total movie duration is $\simeq 83$ hr.  For M87*, where $GMc^{-3} \simeq 9$ hr, the time between frames is $\simeq 2$ days and the total movie duration is $\simeq 15$ yr.  A more detailed description of how the library was made is provided in \cite{SgrAPaperV} and \cite{Wong_2022}.

This paper is organized as follows.  Section \ref{sec:methodology} defines $\Omega_p$ and introduces a methodology for measuring it in idealized synthetic image data. Section \ref{sec:results} applies this method to the Sgr A* model library and discusses the results.  Section \ref{sec:conclusion} provides a summary and describes next steps.  

\section{Measuring Pattern Speed} \label{sec:methodology}

The hot spot model discussed in Section \ref{sec:intro} illustrates the difficulties in defining and measuring rotation in EHT movies.  A single, equatorial, freely orbiting hot spot traces a complicated trajectory on the plane of the sky.  Lensing can produce multiple images.  Lensing has a particularly strong effect when the hot spot is seen edge on in the equatorial plane; then the brightest images trace a trajectory both above and below the black hole shadow.\footnote{The shadow is defined as the region interior to the critical curve, where photon trajectories can be traced back to the event horizon.}  Relativistic foreshortening and lensing make the apparent motion nonuniform; at modest inclination, the hot spot appears to move more quickly as it approaches the observer and more slowly as it recedes.  Clearly there is a lot of potential information in EHT movies.  

In this paper, we set aside this complexity and ask the most basic questions about the motion of brightness fluctuations on the ring.  First, is it possible to determine if the fluctuations circulate clockwise or counterclockwise on the sky?  Second, is it possible to measure a characteristic rotation frequency, or pattern speed, $\Omega_p$?

We begin by reducing the movie data to a manageable form.  In each synthetic image, we sample the surface brightness $T_b$ on a circle defined by 
\begin{eqnarray}\label{eq:ringdef}
x & = & -\sqrt{27} \sin\PA + 2\spin \sin i\\
y & = & \,\,\,\,\sqrt{27} \cos\PA
\end{eqnarray} 
where the position angle ($\PA$) parameterizes the location on the circle, $i$ is the inclination angle between our line of sight and the angular momentum of the disk, and $\spin \equiv Jc/GM^2$ is the dimensionless black hole spin. We use Bardeen's coordinates for $x$ and $y$ expressed in units of $GM/(c^2 D)$ for a distance $D$ to the source. This circle coincides with the critical curve (or shadow boundary) to first order in $\spin$ \citep{Gralla_2020}. 

The synthetic images are smoothed using a Gaussian kernel with FWHM $= 20\,\mu$as, the nominal EHT resolution.  Since there are $\sim 3$ resolution elements across Sgr A*'s ring, the brightness distribution sampled on the ring is insensitive to the precise radius and centering of the circle.  Figure \ref{fig:ringlocation} shows the ring superposed over an example synthetic image.

\begin{figure*}[!t]
    \plottwo{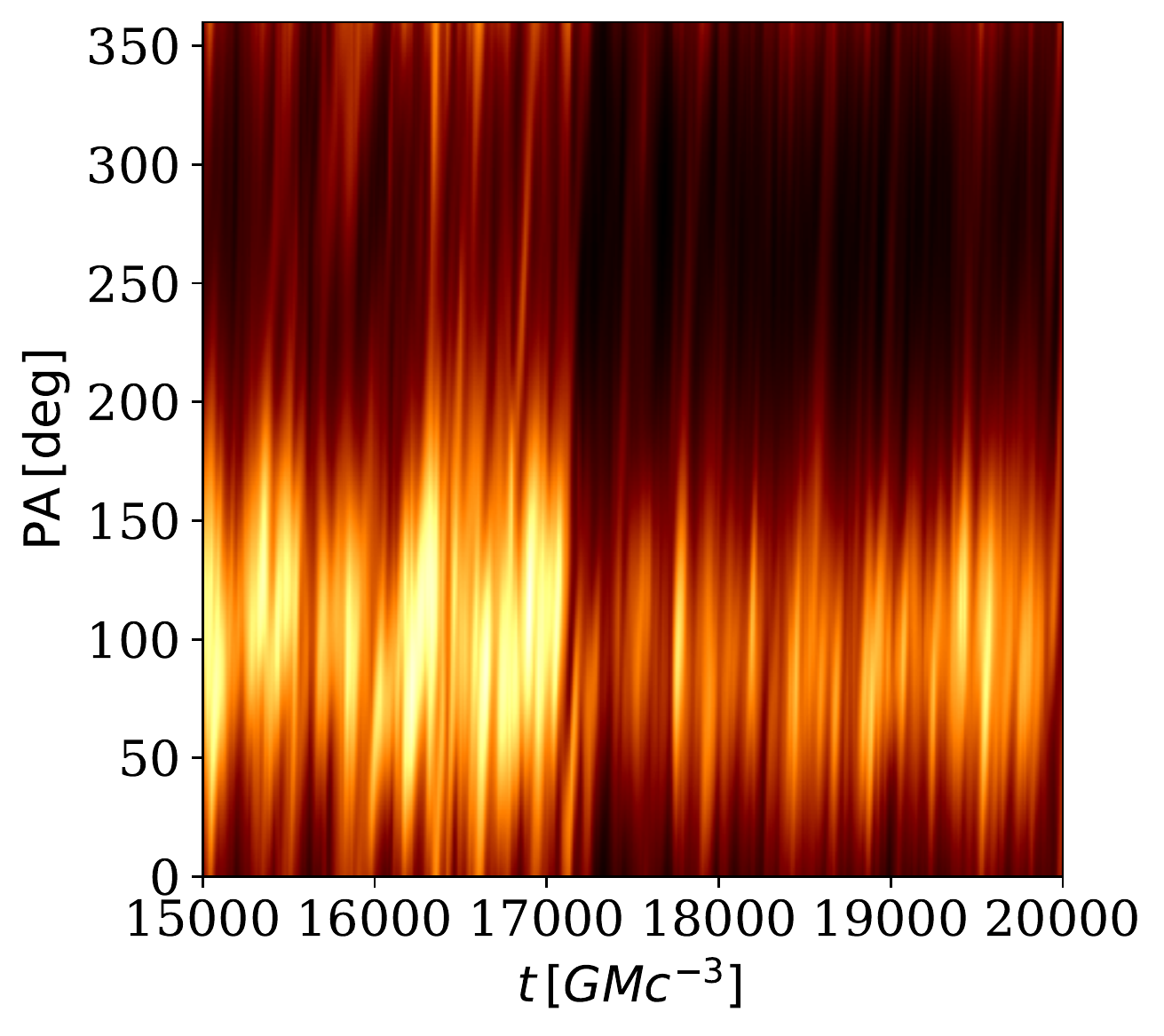}{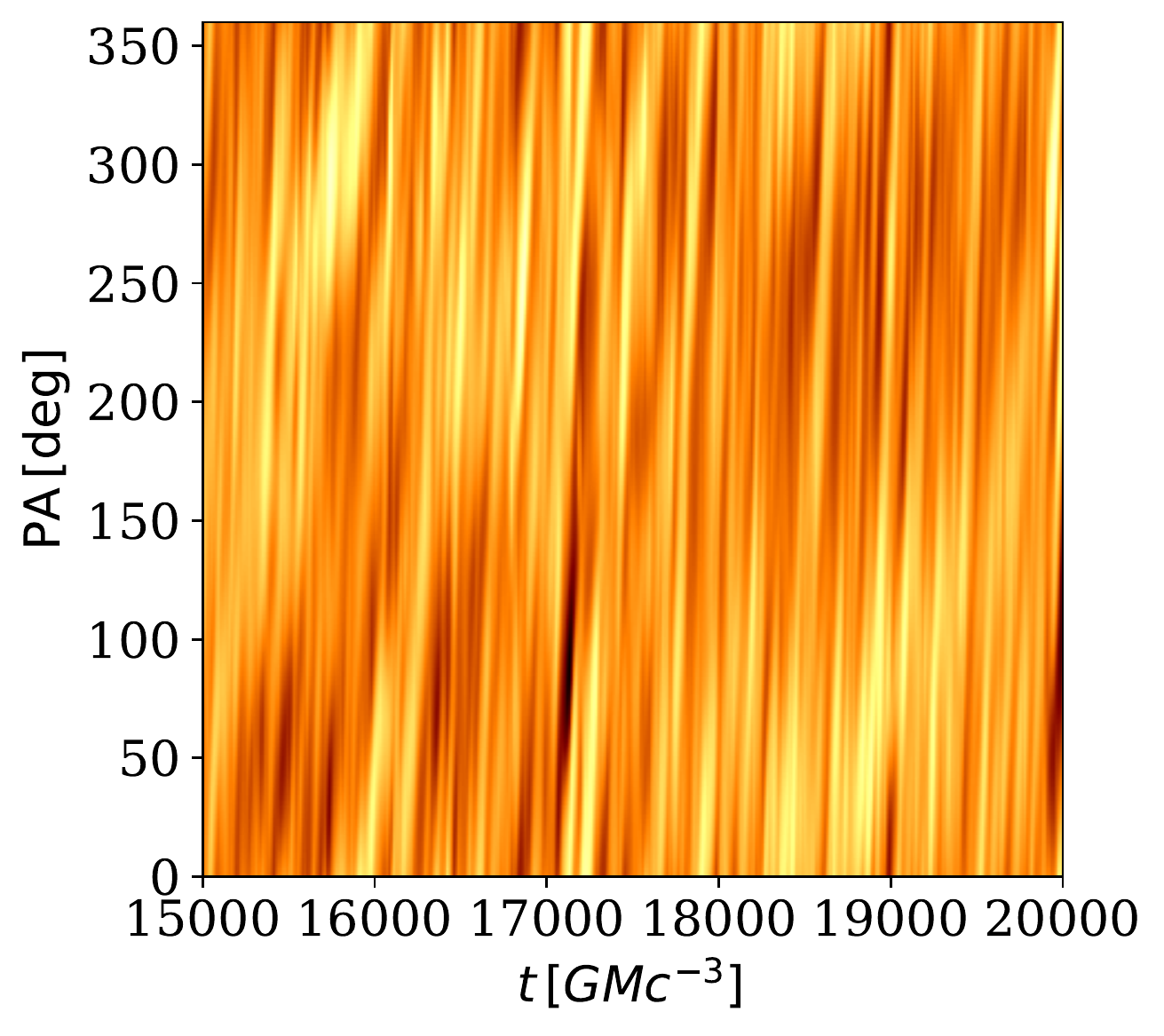}
    \caption{
        Cylinder plot (left) and normalized 
        cylinder plot (right) for the fiducial Sgr A* model (MAD, $\spin = 0.5$, $i = 30^{\circ}$, $R_{high} = 160$). 
        \label{fig:cylinder}
    }
\end{figure*}

Evaluating the surface brightness at each point on the ring over the entire duration of the movie yields $T_b = T_b({\PA}, t)$, which we will call a cylinder plot.  This cylinder plot is periodic in PA.  The left panel of Figure \ref{fig:cylinder} shows a cylinder plot for a fiducial model. Although we sample along a thin ring, blurring to the EHT's nominal resolution causes near-ring bright features to appear on the ring, giving the ring an effective thickness.

The cylinder plot shows characteristic diagonal bands.  These thin bands appear near-vertical simply due to the aspect ratio.  Each band corresponds to the movement of a bright feature around the ring.  The features' orientation implies a net rotation toward positive $\PA$ (counterclockwise on the sky).  The average slope of these features is the pattern speed $\Omega_p$, which we will measure using an autocorrelation analysis of the cylinder plot.  

\subsection{Normalization} 
\label{sec:standard}

The cylinder plot in Figure \ref{fig:cylinder} is (1) brightest at $\PA \approx 90^\circ$, which corresponds to the approaching side of the accretion flow, and (2) exhibits fluctuations in source brightness over time, with a large dip in brightness near $t \simeq 17,500 GMc^{-3}$.  An autocorrelation of the raw cylinder plot will be dominated by a few brightest features and will thus throw away information in low surface brightness features.  

The bright feature in the cylinder plot near $\PA = 90^\circ$ is partially explained by Doppler boosting.\footnote{Flow geometry and lensing also contribute to ring asymmetry.}  This brightness peak would appear even if the emission were axisymmetric.  Both the time-averaged and fluctuating emission are amplified there. The asymmetry dominates the autocorrelation of the cylinder plot, downweighting information from low surface brightness $\mathrm{PAs}$ and reducing the accuracy of $\Omega_p$ measurements. Assuming the signal-to-noise ratio is high, we would like to use the information available from fluctuations at all $\mathrm{PAs}$. 

The source brightness variations likewise amplify both the mean brightness and nonaxisymmetric fluctuations.  Assuming that the signal-to-noise ratio is high, we would like to treat each snapshot on an equal footing. 

In order to weight snapshots and $\mathrm{PAs}$ with different total fluxes more equally, we construct the cylinder plot using $\log(T_b)$. We then normalize by performing a mean subtraction along each time slice and each $\PA$ slice. The resulting cylinder plot has a mean of $0$ along each column and row. This normalization procedure is independent of the order in which the mean subtraction is applied. The mean-subtracted logarithmic cylinder plot produces more accurate pattern speed measurements compared to a mean-subtracted linear plot (see Section \ref{subsec:verification} for accuracy tests). The right panel of Figure \ref{fig:cylinder} shows the normalized cylinder plot, denoted $\tilde{T}_b(\PA, t)$. 

\subsection{From Autocorrelation to Pattern Speed}\label{subsec:moments}

Once the cylinder plot is normalized, we autocorrelate $\tilde{T}_b(\PA, t)$.  Setting $\PA' = \PA + \Delta\PA$ and $t' = t + \Delta t$, the dimensionless autocorrelation function $\xi$ is 
\begin{eqnarray}
\xi(\Delta t, \Delta \PA) & \equiv & \frac{1}{\sigma^2} \left\langle \tilde{T}_b(\PA, t) 
\tilde{T}_b(\PA', t') \right\rangle\\
& = &
\frac{1}{\sigma^2} 
\mathcal{F}^{-1} ( |\mathcal{F}(\tilde{T}_b)|^2 )
\end{eqnarray}
where $\langle\rangle$ denotes an average, $\sigma^2$ is the variance of $\tilde{T}_b$, and $\mathcal{F}$ is the Fourier transform.  Notice that $\tilde{T}_b$ is periodic in $\PA$ but not in $t$.  We do not apply an explicit window function in time.  The discontinuity that results from joining the beginning and ending of the time series with periodic boundary conditions at each $\PA$ produces power at high temporal frequencies that does not affect our analysis.  

\begin{figure}
   \plotone{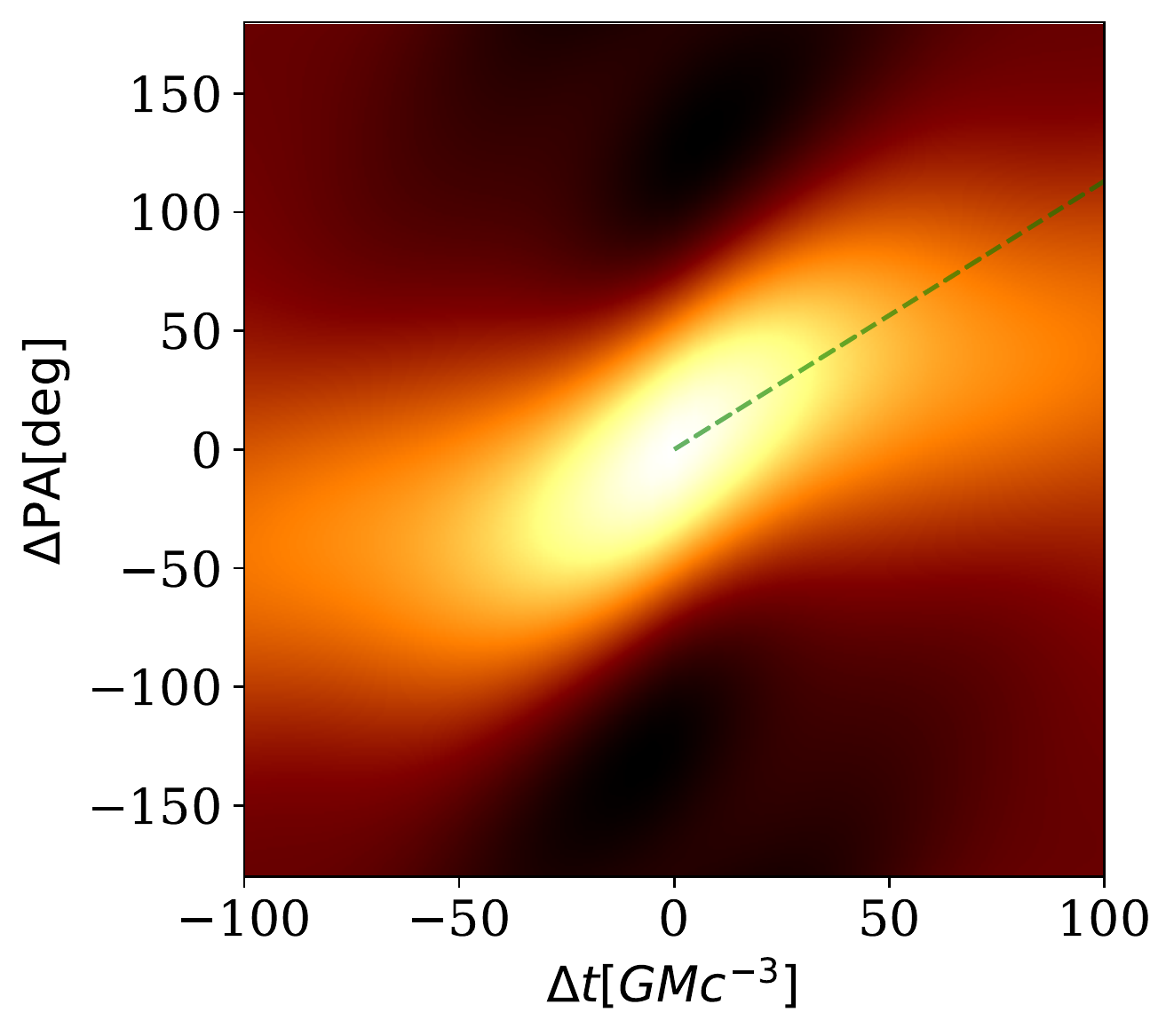}
   \caption{
       Autocorrelation function for the fiducial Sgr A* model.  The correlation function is tilted in the $(\Delta t, \Delta\PA)$ configuration space, and the slope is shown as a green dashed line, corresponding to the pattern speed $\Omega_p$. A positive $\Omega_p$ implies that at $\Delta t > 0$, the fluctuations in surface brightness sampled on the ring are shifted toward positive $\PA$. Notice that the measured slope is drawn through the peaks of the autocorrelation at each $\Delta t$ in the region close to $(\Delta t, \Delta \PA) = (0,0)$.
       \label{fig:RACF1}
   } 
\end{figure}

Figure \ref{fig:RACF1} shows the autocorrelation function for our fiducial model.  Only the central part of the correlation function is shown.  The tilt of the correlation function suggests a pattern speed  $\Omega_p \sim 1^\circ$ per $GMc^{-3}$.   

The pattern speed $\Omega_p$ can be measured using second moments of $\xi$.  In this subsection, for clarity, we use $\phi \equiv \Delta \PA$ and $\tau \equiv \Delta t$ for the arguments of $\xi$.  The relevant second moments are then
\begin{eqnarray}
M_{\tau\tau} &\equiv &  \int \tau^2 \, \xi \, d{\phi} \,d\tau \\
M_{\phi \tau} &\equiv & \int \phi \, \tau \, \xi \, d{\phi} \,d\tau \label{eq:m_phi_t}.
\end{eqnarray}
The domain of integration will be specified below.

We define $\Omega_p$ by applying a shear transformation to the correlation function, and the integration region, until the off-diagonal moment vanishes.  That is, we define $\phi' = \phi - \Omega_p \tau$, and adjust $\Omega_p$ so that $M_{\phi'\tau} = 0$.  Then 
\begin{eqnarray}
    M_{\phi \tau } & = &\int \phi \, \tau \, \xi(\tau, \phi)  \,d\phi \,d\tau  \\
    &= & \int (\phi' + \Omega_p \tau) \, \tau \, \xi(\tau, \phi')  \,d\phi' \,d\tau  \\
    &= & \, M_{\phi'\tau} + \Omega_p \int \, \tau^2 \, \xi(\tau, \phi')  \,d\phi' \,d\tau  \\
    &= & \, \Omega_p \, M_{\tau\tau}.
\end{eqnarray}
The second equality follows from the definition of $\phi'$ (the Jacobian of the shear transformation is $1$; notice that the domain of integration must be transformed as well).  The third equality follows from the definition of the moments.  The final equality follows if $\Omega_p$ is defined so that $M_{\phi'\tau} = 0$.  Thus
\begin{equation}\label{eq:Omega_p}
    \Omega_p = \frac{M_{\phi\tau}}{M_{\tau\tau}},
\end{equation}
which is evidently dimensionally correct.  

The domain of integration should be set to maximize accuracy of the estimate of $\Omega_p$.  A limited number of independent frames is used to estimate $\xi$, which introduces noise in $\xi$. The relative uncertainty in $\xi$ increases away from the origin, and outside a few correlation lengths $\xi$ is completely dominated by noise.  If the domain of integration is too large then the moments are dominated by noise.  Near the origin, however, pixelation of $\xi$ also introduces errors.  If the domain of integration is too small, then accuracy is lost.  With these considerations in mind, we choose to integrate over a region with $\xi > \xi_{\rm crit}$.  We set $\xi_{\rm crit} = 0.8$ to maximize measurement accuracy in our test problems (see Section \ref{subsec:verification}) and minimize outliers in a survey of $\Omega_p$ over the GRMHD model library.  

To summarize, $\Omega_p$ is estimated using the following procedure.  Beginning with a high angular resolution synthetic movie: (1) smooth each frame to the nominal EHT resolution using a Gaussian kernel with FWHM $= 20\,\mu$as; (2) sample the ring specified by Equation~(\ref{eq:ringdef}) in these frames to obtain the cylinder plot $T_b(\PA, t)$; (3) take the log and mean subtract the cylinder plot to obtain $\tilde{T}_b(\PA, t)$ (see Section \ref{sec:standard}); (4) calculate the correlation function $\xi$; (5) evaluate the moments of $\xi$ at $\xi > \xi_{\rm crit}$; and (6) calculate $\Omega_p$ using Equation~(\ref{eq:Omega_p}).\footnote{A copy of the script used to run this procedure is available here: \url{https://doi.org/10.5281/zenodo.7809121} \citep{cylinder_clean}.}

\subsection{Verification}
\label{subsec:verification}

As a first test of the procedure, we created three movies containing a superposition of transient hot spots moving with constant angular frequency of either $\Omega_{\mathrm{hs}} = 1.23, 2.72,$ or $3.14^\circ$ per $GMc^{-3}$ near the photon ring radius. The procedure recovers $\Omega_p = \Omega_{\mathrm{hs}}$ to within $4.3\%$. 

As a second test of the procedure, we produce mock cylinder plots with a known pattern speed using sheared Gaussian random fields $f$.  We begin with an unsheared random field $f_u(\PA, t)$ with a Matern power spectrum $P_u(m, \omega) \propto (1 + (m/m_o)^2 + (\omega \tau_o)^2)^{-5/2}$.  Here $m$ is the angular Fourier coordinate, $\omega$ is the temporal Fourier coordinate, and $m_o$ and $\tau_o$ are constants.  The power spectrum is sheared by setting $P_s = P_u(m, \omega + m \Omega_s)$.  Then a realization of the sheared field is generated in the Fourier domain from $P_s$ and transformed back to a realization in coordinate space $f(\PA, t)$, which is our mock cylinder plot.  Calculating $\Omega_p$ for 500 realizations of $f$ with different shear rates $\Omega_s$, we are able to assess the accuracy of our measurement and the effects of pixelation.  For mock cylinder plot parameters that are similar to those in the model library, we recover $\Omega_p = \Omega_s$ with a root mean squared error of $2.8\%$. 

\section{Pattern Speeds in the Sgr A* Model Library}
\label{sec:results}

The Illinois component of the Sgr A* model library has four parameters:  the magnetic flux (MAD or SANE), the inclination angle $i$, black hole spin $\spin$, and the electron temperature parameter $\rhigh$. In our convention, $i$ is the angle between the line of sight and the accretion flow orbital angular momentum vector. Models with $\spin > 0$ have prograde accretion flows, and models with $\spin < 0$ have counterrotating, or retrograde, accretion flows. All models are assumed to have spin parallel or antiparallel to the accretion flow angular momentum; they are untilted. The electron temperature is set using the $\rhigh$ model, in which the ratio of the ion-to-electron temperature varies smoothly from $1$ where $\beta \ll 1$ to $\rhigh$ where $\beta \gg 1$. For more details on this prescription, see \cite{Wong_2022}, Equation~(22). We have measured $\Omega_p$ across the entire model library.  Table \ref{tab:long} in the appendix lists $\Omega_p$ for all models.

\subsection{Sub-Keplerian Pattern Speeds}
\label{sec:subkepler}

Our first main finding is that $\Omega_p$ is small compared to what one would expect in a Keplerian hot spot model.  The largest value measured in the entire library is $2.60^\circ$ per $GMc^{-3}$, and a more typical value is $1^\circ$ per $GMc^{-3}$ (see Table \ref{tab:typical_values}).  Thus $\Omega_p \simeq \Omega_K(r = 4\,GMc^{-2})/7$. This is small compared to what one would expect for hot spots orbiting freely close to the radius of peak emission. 
\begin{deluxetable}{cccc}
\tablecaption{Typical Pattern Speeds from the Sgr A* Library \label{tab:typical_values} }
\tablehead{\colhead{MAD/SANE} & \colhead{$i$} &  \colhead{$|\overline{{\Omega}_p}|$ ($^\circ/GMc^{-3}$)} & \colhead{STD ($^\circ/GMc^{-3}$)}  }
\startdata
All & All & $0.72$ & $0.47$ \\
All & Face on & $1.13$ & $0.43$ \\
MAD & All & $0.76$ & $0.32$ \\
MAD & Face on & $1.04$ & $0.16$ \\
SANE & All & $0.68$ & $0.57$ \\
SANE & Face on & $1.22$ & $0.58$
\enddata
\tablecomments{Mean values of $|\Omega_p|$ and their standard deviations, averaged over varying magnetic flux types (MAD, SANE, or both) and inclinations ($i \in \{10^\circ, 170^\circ \}$ for face-on models, or $i \in \{10^\circ, 30^\circ, 50^\circ \dots\, 170^\circ\}$ for all models). Inclination dominates the standard deviation when averaging over all inclinations; spin dominates the standard deviation for face on only. SANEs have a larger standard deviation than MADs.}
\end{deluxetable}

\begin{figure*}
    \plotone{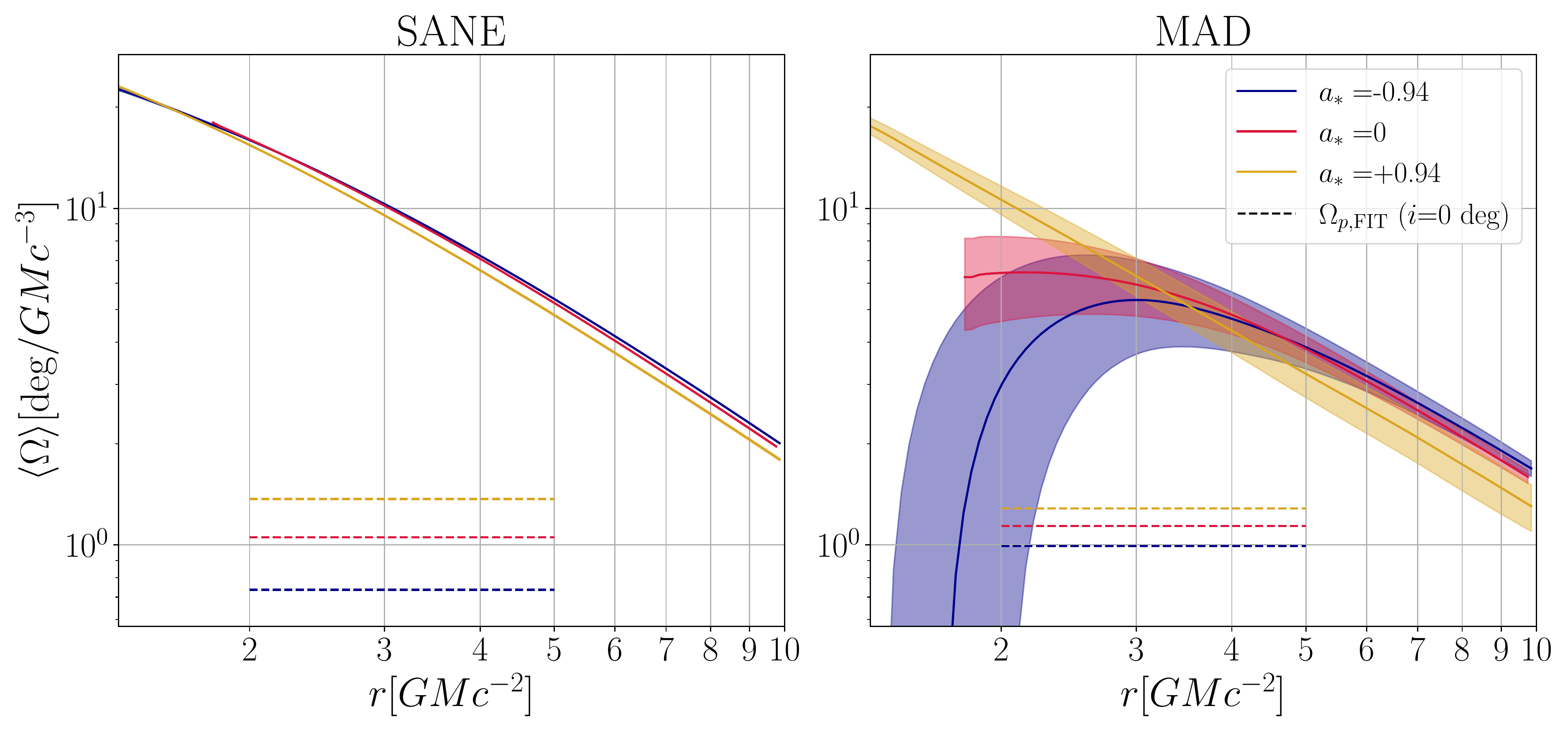}
    \caption{ Mean orbital frequency of the underlying fluid averaged over time and azimuth, for SANEs (left) and MADs (right), shown as solid lines. The bands around each orbital frequency designate 1 standard deviation. The SANE orbital frequencies are very nearly the Keplerian (circular orbit) value and have too small a standard deviation band to be seen. The best-fit pattern speeds (for $i = 0^\circ$, averaged across $\rhigh$, measured at the critical curve angular radius) from Equations \ref{eq:fitSANE} and \ref{eq:fitMAD} are shown as dashed lines over the primary emitting region. Color shows spin. The pattern speed is slower than the azimuthal frequency of the fluid and the Keplerian frequency.
        \label{fig:omega_fluid}
    }
\end{figure*}

The plasma is not orbiting freely, however, and instead exhibits pressure-driven and magnetic field-driven velocity fluctuations. In MAD models, the azimuthal fluid velocity is strongly sub-Keplerian. Could this explain the low pattern speeds? Figure \ref{fig:omega_fluid} shows the orbital frequency of the underlying plasma $\langle u^\phi\rangle/\langle u^t\rangle$, as seen by a distant observer in spherical Kerr–Schild coordinates. The figure shows the time- and longitude-averaged mean for both MADs and SANEs with a band indicating one standard deviation around the mean. The measured pattern speeds, which are shown as dashed lines spanning the principal emission region, are well below the fluid velocity. In SANE models, the azimuthal fluid velocity is indistinguishable from Keplerian.  Apparently sub-Keplerian azimuthal fluid velocities do not provide a consistent explanation for the low pattern speeds.   

\begin{figure*}
    \plottwo{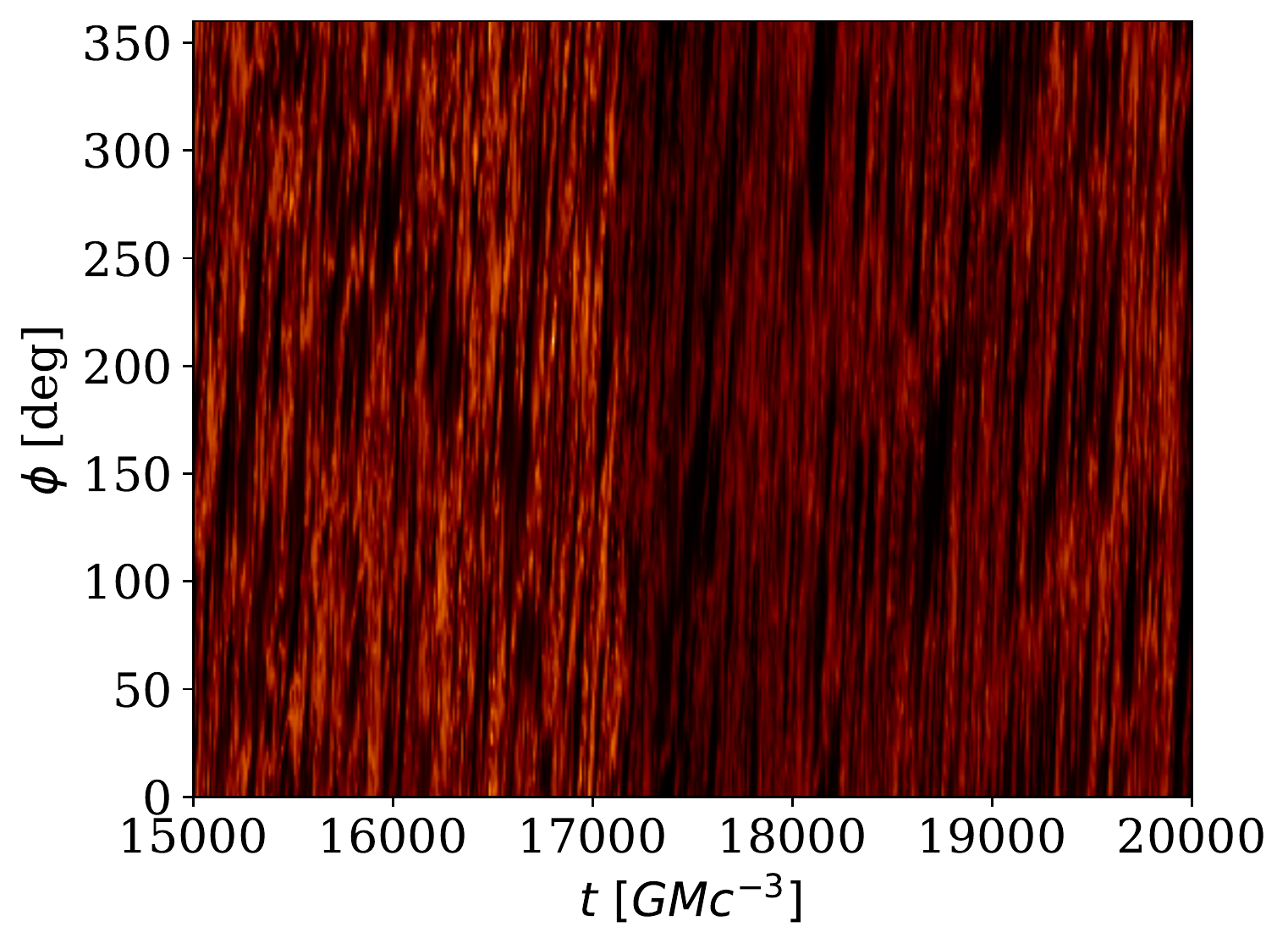}{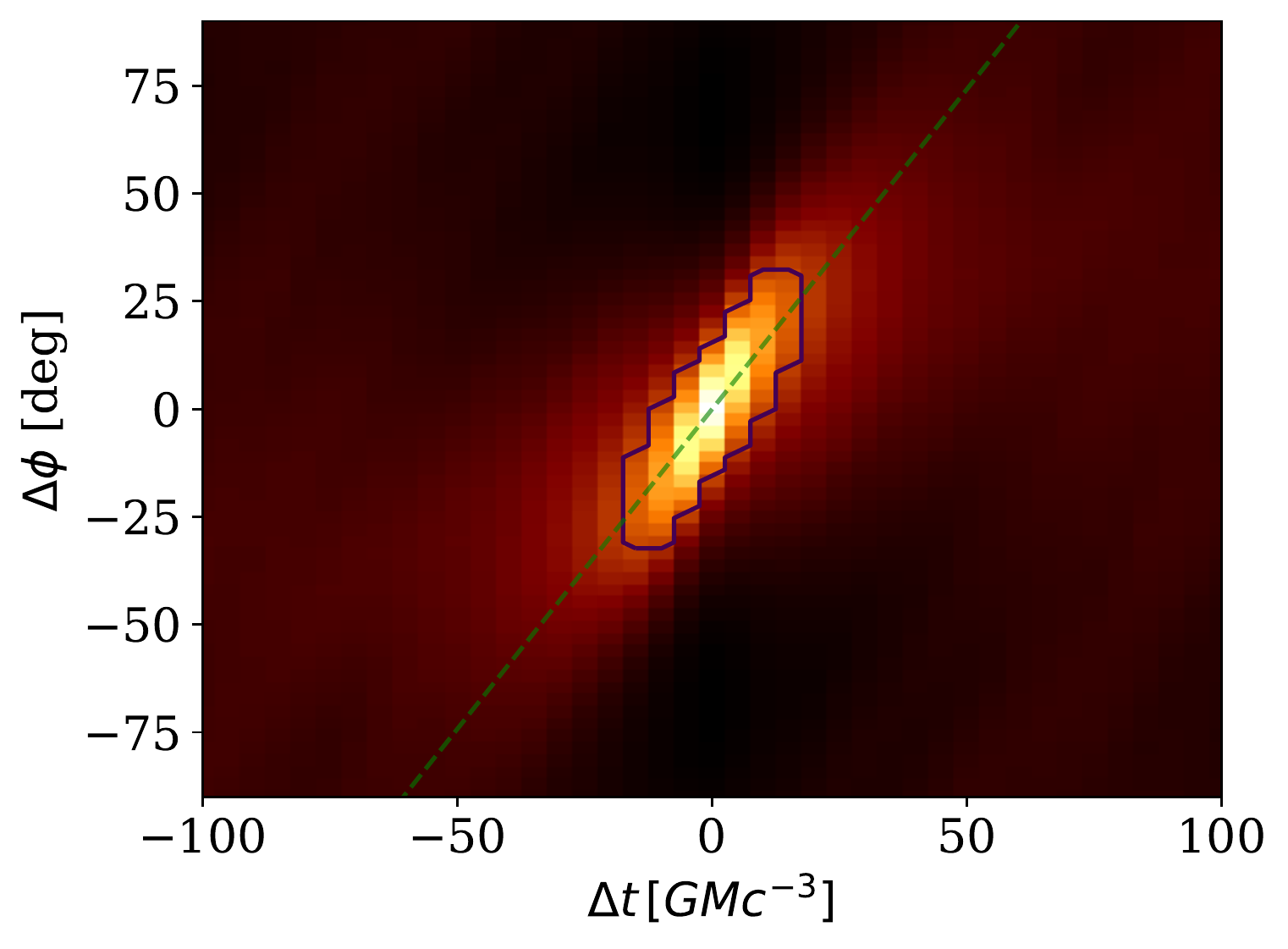}
    \caption{
        An unnormalized cylinder plot for the gas pressure at the midplane at $r = 3\,GMc^{-2}$ (left) in our fiducial GRMHD simulation (MAD, $\spin = 0.5$), and its autocorrelation function $\xi$ (right), where the blue line shows $\Omega_p = 1.5^\circ$ per $GMc^{-3}$.  This pattern speed is measured from the region with $\xi > 0.3$, which is inside the black contour.  The vertical coordinate $\phi$ is the longitude in spherical Kerr–Schild coordinates.
        \label{fig:pgcylinder}
    }
\end{figure*}

The pattern speed seen in movies is, however, similar to the pattern speed measured in gas pressure ($p_g$) fluctuations. We measured $p_g(r = 3\,GMc^{-2}, \theta = \pi/2, \phi, 1.5 \times 10^4 < t c^3/(G M) < 3 \times 10^4)$ to make a cylinder plot of gas pressure in the fiducial model.  Figure \ref{fig:pgcylinder} shows the raw cylinder plot and, after normalization, the correlation function.  The pattern speed for the gas pressure fluctuations is $\Omega_p = 1.1^\circ$ per $GMc^{-3}$, while the pattern speed for the synthetic images is $\Omega_p = 1.1^\circ$ per $GMc^{-3}$.  We find similar results in other models. It seems the sub-Keplerian pattern speeds are not an artifact of the low angular resolution in the images, radiative transfer, or lensing: they result from sub-Keplerian pattern speeds in the accretion flow itself.

\begin{figure*}
    \plotone{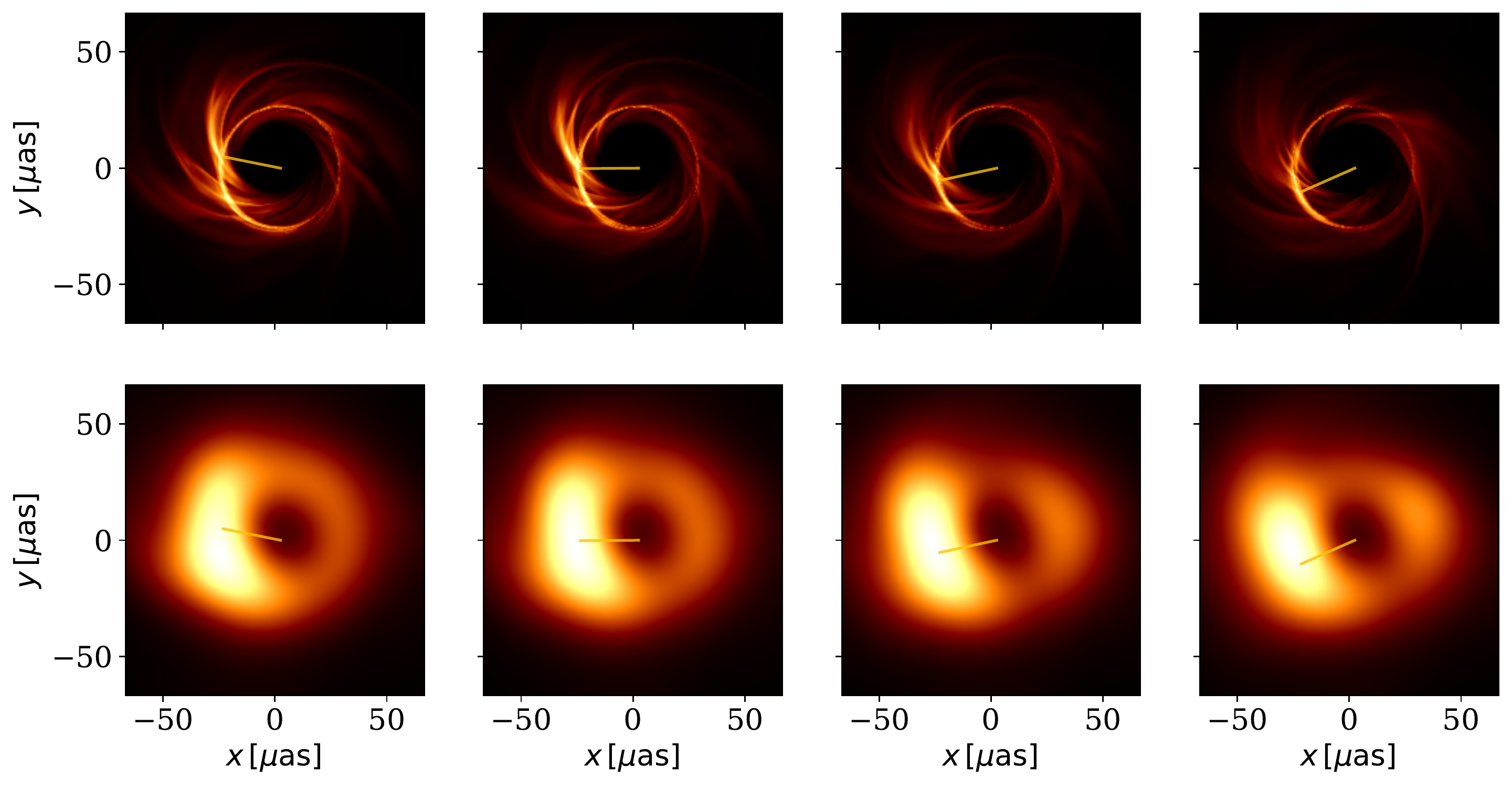}
    \caption{Four frames from an animation of the unblurred (top) and blurred (bottom) Sgr A* fiducial model, with a clock arm rotating at the measured pattern speed at the critical curve angular radius.  The example frames are separated at a cadence of $10\,GMc^{-3}$.  In the online animation, we show an unblurred (left) and blurred (right) movie of the Sgr A* fiducial model from $15,000$ to $20,000 \, GMc^{-3}$.  By eye, we can see spiral shocks moving around the ring at the mean measured pattern speed, with some variation about the mean. The Keplerian velocity is over $5 \times$ the pattern speed, which is clearly too fast to fit the mean angular phase velocity of these spiral shocks. The real-time duration of the animation is 100 s. (An \href{https://iopscience.iop.org/article/10.3847/1538-4357/acd2c8}{animation} of this figure is available.)
        \label{fig:spinner}    }
\end{figure*}

The online animated Figure \ref{fig:spinner} shows the evolution of the image at both the full resolution of our synthetic images and the nominal EHT resolution.  The clock hand in the animation moves at the pattern speed $\Omega_p$ for this model, measured on the ring in the blurred images.  The full resolution animation shows that the pattern speed is tracking narrow, trailing spiral features.  The spiral features, like nearly all emission in MAD models, arise close to the midplane (see Figure 4 of \citealt{M87PaperV}).  

The pattern speed associated with brightness fluctuations in the images and gas pressure fluctuations in the GRMHD models is slow compared to both the Keplerian speed and the azimuthal speed of the plasma.  It must therefore be measuring a wave speed in the plasma.  Since we see emission from only a narrow band in radius in these images 
(see the images in Figure \ref{fig:spinner} and the emission map in Figure 4 of \citealt{M87PaperV}), we must be seeing the azimuthal phase speed of the wave.  

The underlying wave field is a combination of linear and nonlinear (shock) excitations.  For simplicity, let us adopt a linear, hydrodynamic model, with a wave $\propto \exp(i k r + i m\phi - i \omega t)$.  In the tight winding (WKB) approximation, assuming that the disk is circular and Newtonian with orbital frequency $\Omega$, the well-known dispersion relation is $(\omega - m\Omega)^2 = \Omega^2 + c_s^2 k^2$.  Here $c_s$ is the sound speed.  The azimuthal component of the phase velocity is then $\omega/m = \Omega \pm (\Omega^2 + c_s^2 k^2)^{1/2}/m$.  The phase velocity can thus be made small for the negative root and an appropriate choice of $k$ and $m$.  If the wave is trailing, as seen in the simulations, then the low phase velocity waves are ingoing. A nonlinear version of this argument is presented in \cite{Spruit_1987}, which demonstrates the existence of stationary (zero pattern speed!) shocks in Keplerian disks.  

Ingoing waves are plausibly excited by turbulence at larger radii.  The pattern speed would correspond to the orbital frequency of the underlying plasma at the excitation radius (i.e. the corotation radius, where each dashed line of Figure \ref{fig:omega_fluid} would intersect with the corresponding solid line). For $\spin = 0$, the corotation radius $\sim 7^{2/3} \times  4\,GMc^{-2} \simeq 15\,GMc^{-2}$, well outside the region currently visible to the EHT. This suggests that disk fluctuations at radii outside the main emission region are important for determining variability. 

\subsection{Parameter Dependence}
\begin{figure*}
    \plottwo{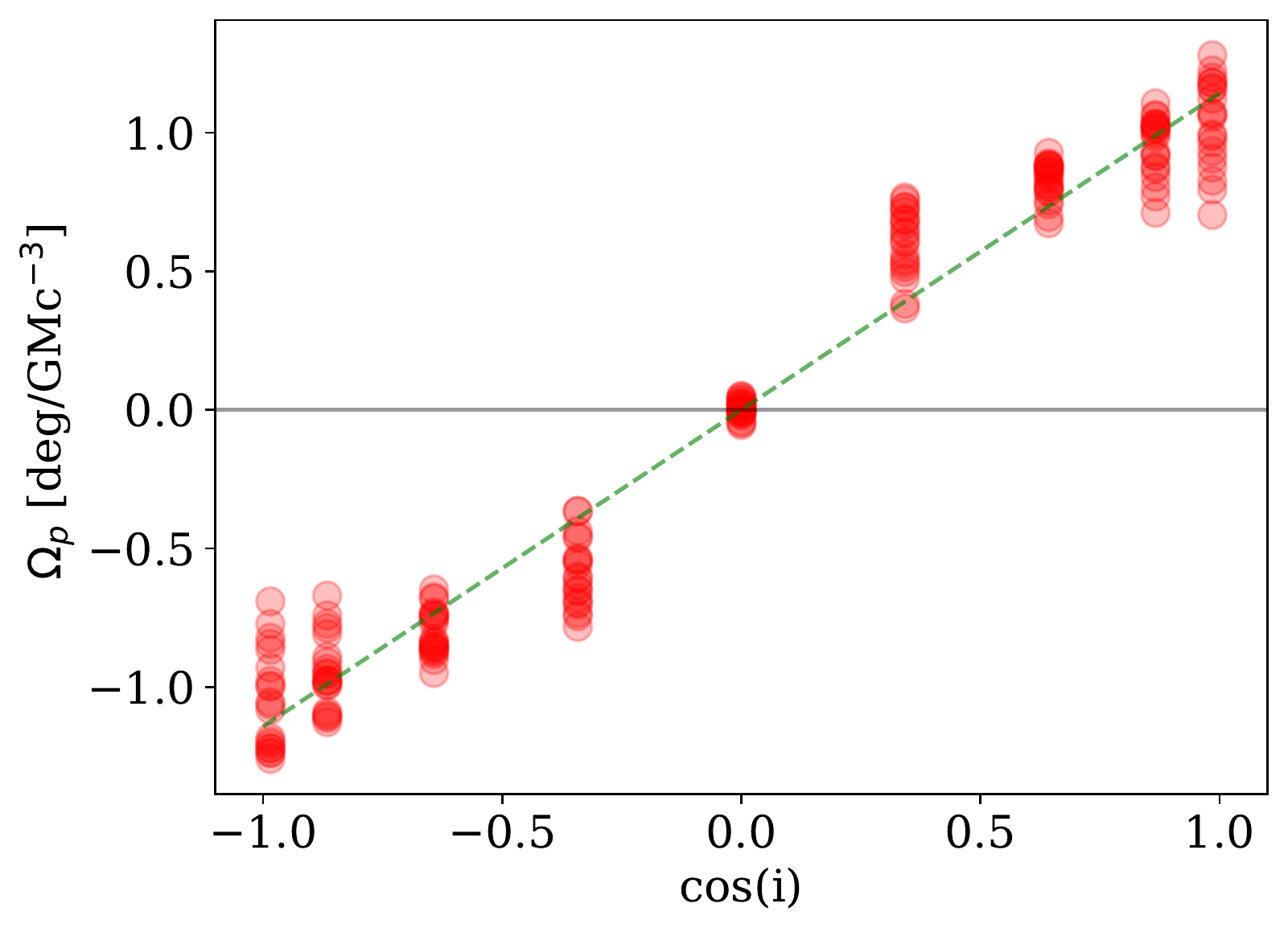}{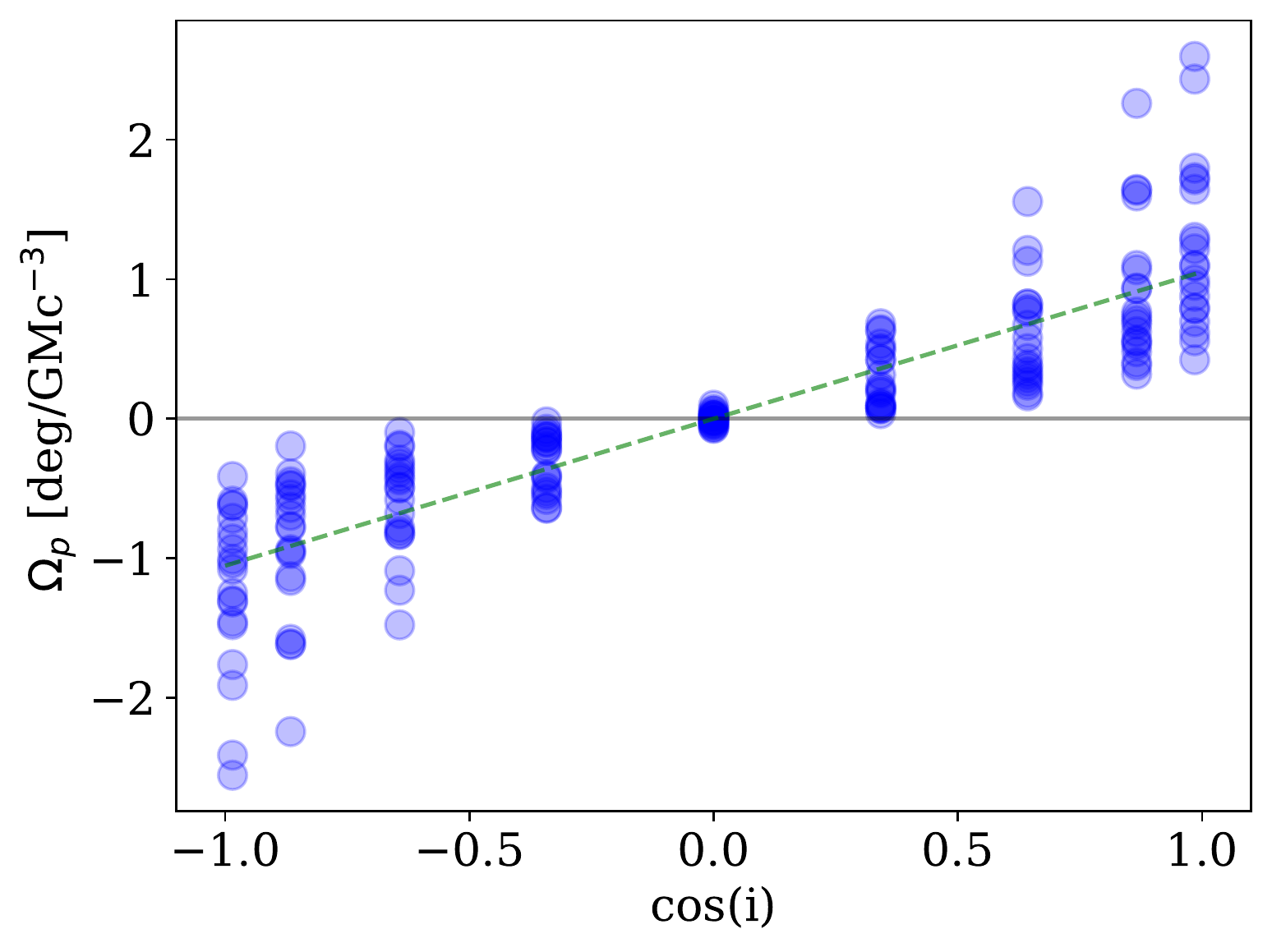}
    \caption{
        Measured pattern speed $\Omega_p$ from the model library plotted against $\cos i$ for MAD (left) and SANE (right) models.  The lines show a linear fit neglecting any dependence on spin and $\rhigh$. Notice the line underestimates $\Omega_p$ in MAD models of $i = \{70^\circ, 110^\circ\}$.
        \label{fig:inclination}
    }
\end{figure*}

Our second main finding is that $\Omega_p$ changes sign from $i < 90^\circ$ to $i > 90^\circ$ (recall that $i$ is the angle between the line of sight and the accretion flow orbital angular momentum vector), so that the sign of $\Omega_p$ signals the sense of rotation projected onto the sky.  This stands in contrast to the time-averaged ring asymmetry, which follows the sign of $\spin$ \citep{M87PaperV}.  It is therefore possible to distinguish between prograde and retrograde accretion in M87* by measuring both the ring asymmetry and $\Omega_p$ (see \citealt{Ricarte_2022} for another technique for distinguishing retrograde accretion). This assumes the Sgr A* models considered here and M87* models have similar $\Omega_p$ in degrees per $GMc^{-3}$, which is what we find in a sparse sampling of the M87* model library. Since the observed asymmetry shows that the spin vector in M87* is pointed away from Earth \citep{M87PaperV}, $\Omega_p < 0$ would imply prograde accretion and $\Omega_p > 0$ retrograde accretion.

Our third main set of findings concerns the dependence of $\Omega_p$ on the model parameters. Fitting to Table \ref{tab:long} (in Section \ref{subsec:appendix}), we find
\begin{equation}\label{eq:fitSANE}
\Omega_p \approx (1.5 + 0.4\,\spin - 0.2 
\ln \rhigh) \cos i, \qquad\mathrm{SANE}
\end{equation}
\begin{equation}\label{eq:fitMAD}
\Omega_p \approx (1.2 + 0.2\,\spin) \cos i, \qquad\mathrm{MAD}
\end{equation}
in units of degrees per $GMc^{-3}$. The SANE fits have a root mean squared error of $0.31^\circ$ per $GMc^{-3}$, while the MAD fits have a root mean squared error of $0.14^\circ$ per $GMc^{-3}$. Systematic errors are discussed in Section \ref{subsec:slowlight}. The worst-fitting models are generally SANE with $\spin > 0$ and low $\rhigh$.  These models tend to have larger $\Omega_p$ than otherwise expected. In Figure \ref{fig:inclination}, we show the measured pattern speeds and the fits from Equations~\ref{eq:fitMAD} and~\ref{eq:fitSANE}. The maximum SANE error is $1.15^\circ$ per $GMc^{-3}$ larger than predicted, while the maximum MAD error is $0.42^\circ$ per $GMc^{-3}$ larger.

The fits in Equations~(\ref{eq:fitSANE}) and (\ref{eq:fitMAD}) are consistent with the relatively slow rotation rate noted above ($\sim 1^\circ$ per $GMc^{-3}$) and with the sign of $\Omega_p$ following the inclination ($\cos i$ dependence).  Notice that for particles moving on a ring with angular frequency $\Omega$, in flat space, the time-averaged apparent rotation rate would be $\Omega \,\mathrm{sgn} (\cos i)$.  

The strongest dependences are on the inclination and mass (the mass dependence is in the units of Equations \ref{eq:fitSANE} and \ref{eq:fitMAD}).  In Sgr A*, the mass is known accurately from independent measurements \citep{Schoedel_2002, Ghez_2003, Ghez_2008, Do_2019, GRAVITY_2019, GRAVITY_2020_precession, SgrAPaperIV}. A measurement of $\Omega_p$ would thus constrain the inclination.  In M87*, if we assume that inclination is determined by the large-scale jet, then $|\cos i| \simeq 1$, so a measurement of $\Omega_p$ would provide a distance-independent constraint on the mass.

The spin dependence of $\Omega_p$ is weak but nonzero. It would seem to require accurate model predictions, lengthy observed movies, and careful interpretation of sparse interferometric data to use this dependence to constrain the spin.

We can estimate the uncertainty of these inclination, mass, and spin constraints using a probability distribution for $\Omega_p$ at each set of parameter values obtained from kernel density estimation. This incorporates the uncertainties in measuring $\Omega_p$ in movies of duration comparable to the expected observations (see Sections \ref{subsec:verification}, \ref{subsec:long_timescale_variability}, and \ref{subsec:observing}). A single $\Omega_p$ measurement could constrain inclination with a standard deviation of $\sim 20^{\circ}$, with slightly smaller errors when the source is edge on and slightly larger errors when the source is face on. For M87*, if we assume the angle of the large-scale jet aligns with the inclination of the accretion flow, then an $\Omega_p$ measurement would provide a distance-independent mass constraint with $1 \sigma$ error of $\sim 33\%$. 

The spin is unconstrained by a single $\Omega_p$ measurement. Instead, the sign of $\Omega_p$ will determine whether $i > 90^{\circ}$ or $i < 90^{\circ}$. From there, the location of the peak brightness temperature will reveal whether the accretion is prograde or retrograde \citep[see Figure 5 of][]{M87PaperV}. The spin is better constrained by making multiple $\Omega_p$ measurements across different radii (see the right panel of Figure \ref{fig:resolution_radius}). These uncertainties do not account for the systematic errors in our models (see, e.g., Section \ref{subsec:slowlight}), uncertainty in movie reconstructions from incomplete $(u,v)$ coverage, or more informative priors.

Finally, notice that $\Omega_p$ is nearly independent of $\rhigh$. This suggests but does not prove that measurements of $\Omega_p$ are insensitive to electron temperature assignment schemes. $\Omega_p$ has a stronger $\rhigh$ dependence in SANE models, plausibly due to the stronger $\rhigh$ dependence of the emission region latitude in SANEs compared to MADs \citep[see Figure 4 of][]{M87PaperV}. This limited $\rhigh$ dependence is also consistent with the close connection between the pattern speed in the gas pressure and the pattern speed in the images noted in Section \ref{sec:subkepler}.

\subsection{Dependence on Resolution and Sampling Radius}

\begin{figure*}
    \centering
    \plottwo{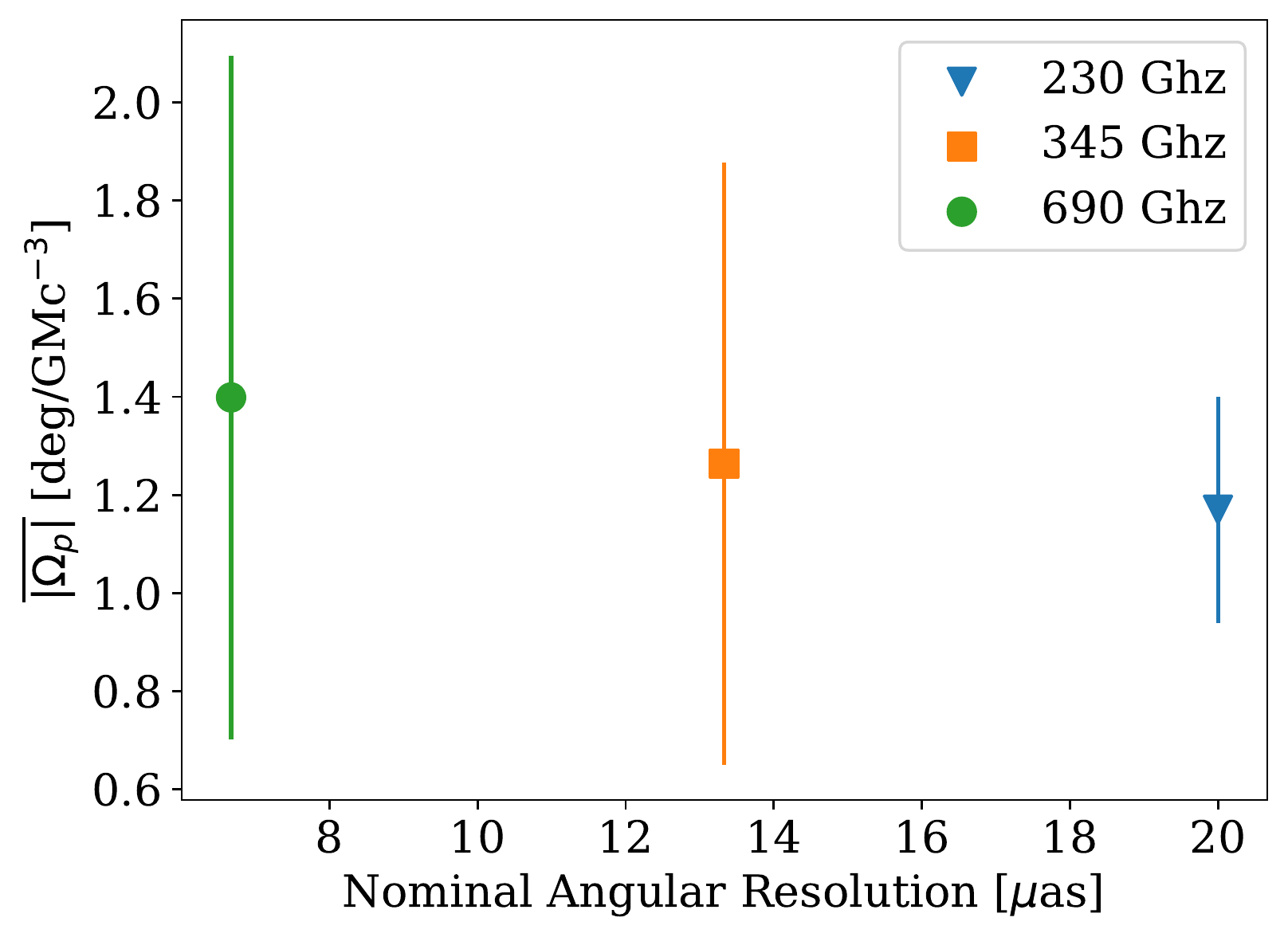}{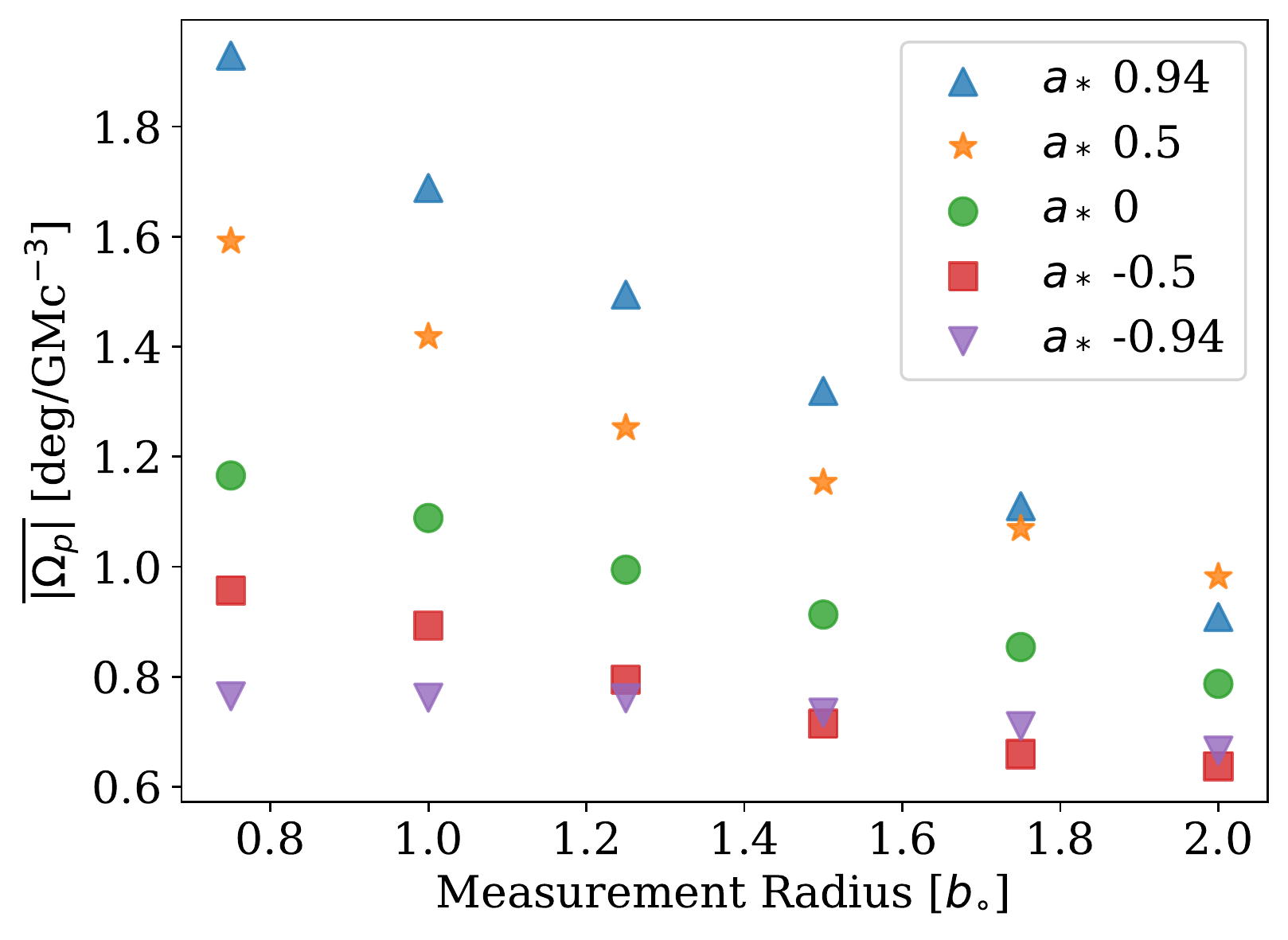}
    \caption{
        The magnitude of ${\Omega}_p$ averaged across all face-on models ($i = 10^\circ, 170^\circ$) from time $15\times 10^3$ to $20 \times 10^3\,GMc^{-3}$, shown at different angular resolutions (left) and different radii (right).  On the left, the nominal angular resolution corresponds to the FWHM of the Gaussian kernel used to smooth the synthetic images.  The blue triangle corresponds to an angular resolution equivalent to 230 GHz observations, where the majority of this analysis was performed. The orange square and the green circle correspond to 345 and 690 GHz observations respectively, which exhibit small increases in $\Omega_p$ and substantial increases in the variation over the model parameters, shown here as one standard deviation error bars measured over all MAD models. On the right, much of this analysis was done at the apparent critical curve angular radius $b_o$; measurements at higher radii lead to a smaller average ${\Omega}_p$.  The blue upward triangle, orange star, green circle, red box, and purple downward triangle correspond to $\spin = 0.94, 0.5, 0, -0.5,$ and $-0.94$ respectively.
        \label{fig:resolution_radius}
     }
\end{figure*}

The EHT may observe at 345 GHz, providing higher resolution than existing data, which are taken at 230 GHz.  Long-baseline space VLBI observations may enable even higher resolution.  It is natural to ask whether the measured pattern speed changes with angular resolution.  We can assess this by revisiting our analysis and smoothing the movies to different resolutions, recalling that up to now we have used Gaussian smoothing with FWHM $= 20\,\mu$as, appropriate to the EHT's nominal resolution at 230 GHz.  The left panel in Figure \ref{fig:resolution_radius} shows the dependence of $|\Omega_p|$ averaged across face-on Sgr A* models.  It seems $\Omega_p$ is only weakly dependent on resolution.
 
Our standard procedure for measuring $\Omega_p$ samples brightness temperature fluctuations on a circle with angular radius $b_o = \sqrt{27} G M/(c^2 D)$, per Equation~(\ref{eq:ringdef}).  This is close to the critical curve where the source is brightest and fluctuations are easiest to measure.  What would happen if fluctuations were measured at other impact parameters?  
The resulting ``rotation curve'' for nearly face-on models is shown in the right panel of Figure \ref{fig:resolution_radius}.  When averaging over all  inclinations, we find a fit consistent with $\Omega_p \simeq 0.7 (b_o/\sqrt{27})^{-1/2}$, although the fit only covers a factor of $4$ in $b_o$, so the scaling is not strongly constrained.  Evidently nonaxisymmetric fluctuations are not dominated by a single pattern speed at all impact parameters; instead there is spectrum of fluctuations with the dominant $\Omega_p$ varying with $b_o$. The impact parameter dependence also changes with spin, as seen in the right panel of Figure \ref{fig:resolution_radius}. Positive-spin (prograde) models show a greater change in $\Omega_p$ with radius than negative-spin (retrograde) models. 

\subsection{Long-timescale Variability}
\label{subsec:long_timescale_variability}

We have checked for consistency of $\Omega_p$ over time by subdividing our movies into three subintervals of duration $5\times 10^3 \,GMc^{-3}$ and measuring $\Omega_p$ in each one.  We find that the standard deviation between subintervals, averaged over all models, is $\sim 0.1^\circ$ per $GMc^{-3}$. Analysis of shorter-duration subintervals can be found in Section \ref{subsec:observing}.

The variation between subintervals is larger for SANE models than MAD models, and for models with $\spin = 0.94$.  The variation is smaller for models with $\rhigh = 1$.  This long-timescale sample variance sets a fundamental limit on how accurately $\Omega_p$ can be measured.

\subsection{Light, Fast and Slow}
\label{subsec:slowlight}

\begin{figure*}[!t]
    \plottwo{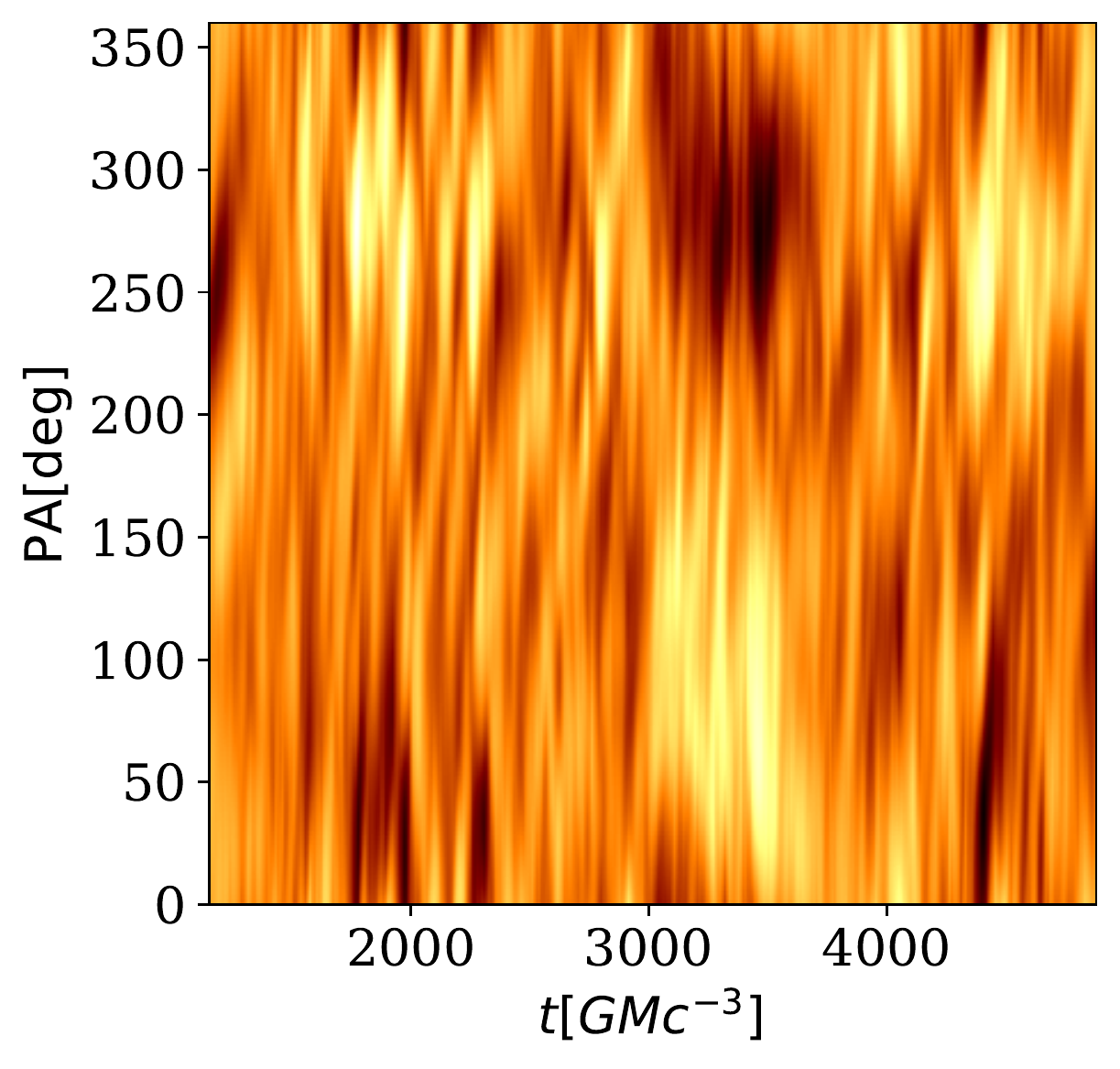}{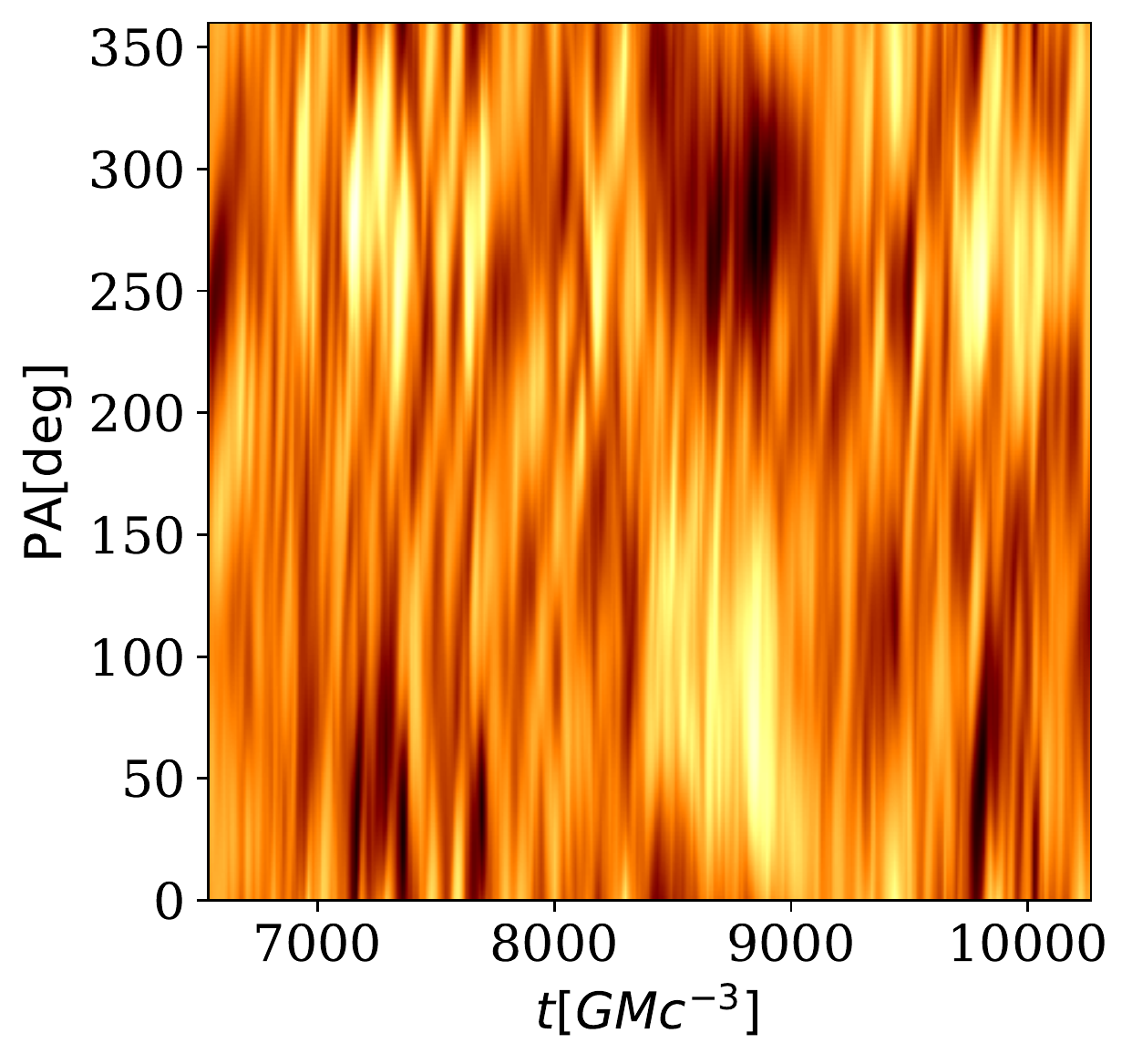}
    \caption{
        Normalized
        cylinder plots in fast light (left) and slow light (right). The image cadence for both is $0.5 \mathrm{GMc^{-3}}$. This particular model is MAD, $\spin = 0.94$, $i = 60^{\circ}$, $\rhigh = 40$.  
        \label{fig:slowlight}
    }
\end{figure*}

Our model images are generated using the ``fast light'' approximation, which freezes the model on a single time slice and then ray traces through that time slice.  Fast light is used because it is simple and the code is easy to parallelize.  However, the fast light approximation fails to accurately represent changes in the source that occur on the light-crossing time.  Short-timescale variations can be accurately captured by ray-tracing through an evolving GRMHD model, a procedure known as ``slow light.''  Does the fast light approximation compromise our estimates of $\Omega_p$?
 
Figure \ref{fig:slowlight} shows normalized cylinder plots for the fast light and slow light versions of a moderately inclined, high-spin model where fast light might be expected to have difficulty (MAD, $\spin = 0.94$, $i = 60^{\circ}$, $\rhigh = 40$).  The cylinder plots differ in detail, especially on short timescales: we find the fast light model has $\Omega_p = 0.31^\circ$ per $GMc^{-3}$, while the slow light model has $\Omega_p = 0.37^\circ$ per $GMc^{-3}$, an increase of $0.06^\circ$ per $GMc^{-3}$ or $19\%$.  This increase is not large enough to change our conclusions, but it does imply additional uncertainty in Equations~(\ref{eq:fitSANE}) and (\ref{eq:fitMAD}) that cannot be accurately evaluated without a slow light model library.

\subsection{Pattern Speed Measurements in Observations} 
\label{subsec:observing}

Measurements of $\Omega_p$ may be affected by the observing cadence, duration, and by limited $(u,v)$ coverage. To check the effect of increased cadence, we have measured $\Omega_p$ in the fast light model used above (MAD, $\spin = 0.94$. $i = 60^{\circ}$, $\rhigh = 40$) at $0.5, 1, 2, 4,$ and $8\,GMc^{-3}$, finding $\Omega_p = 0.31, 0.31, 0.31, 0.30,$ and $0.26^\circ$ per $GMc^{-3}$ respectively.
We have also measured $\Omega_p$ in the example model from Section \ref{sec:methodology} (MAD, $\spin = 0.5$, $i = 30^{\circ}$, $R_{high} = 160$) at a cadence of $5, 10,$ and $20 \, GMc^{-3}$, finding $\Omega_p = 1.1, 1.1,$ and $0^\circ$ per $GMc^{-3}$ respectively. 

For both models, $\Omega_p$ decreases with the cadence, as the fastest features are often short lived. $\Omega_p$ is nearly independent of the cadence below a threshold of $10\,GMc^{-3}$. Beyond this threshold, the autocorrelation peak begins to drop off more steeply, and pixelation effects limit our accuracy (see the end of Section \ref{subsec:moments}). Lowering $\xi_{\rm crit}$ from 0.8 to 0.4, we find $\Omega_p = 0.94$, a 25\% decrease, for a $20 \, GMc^{-3}$ cadence.

\begin{figure*}[!t]
    \plottwo{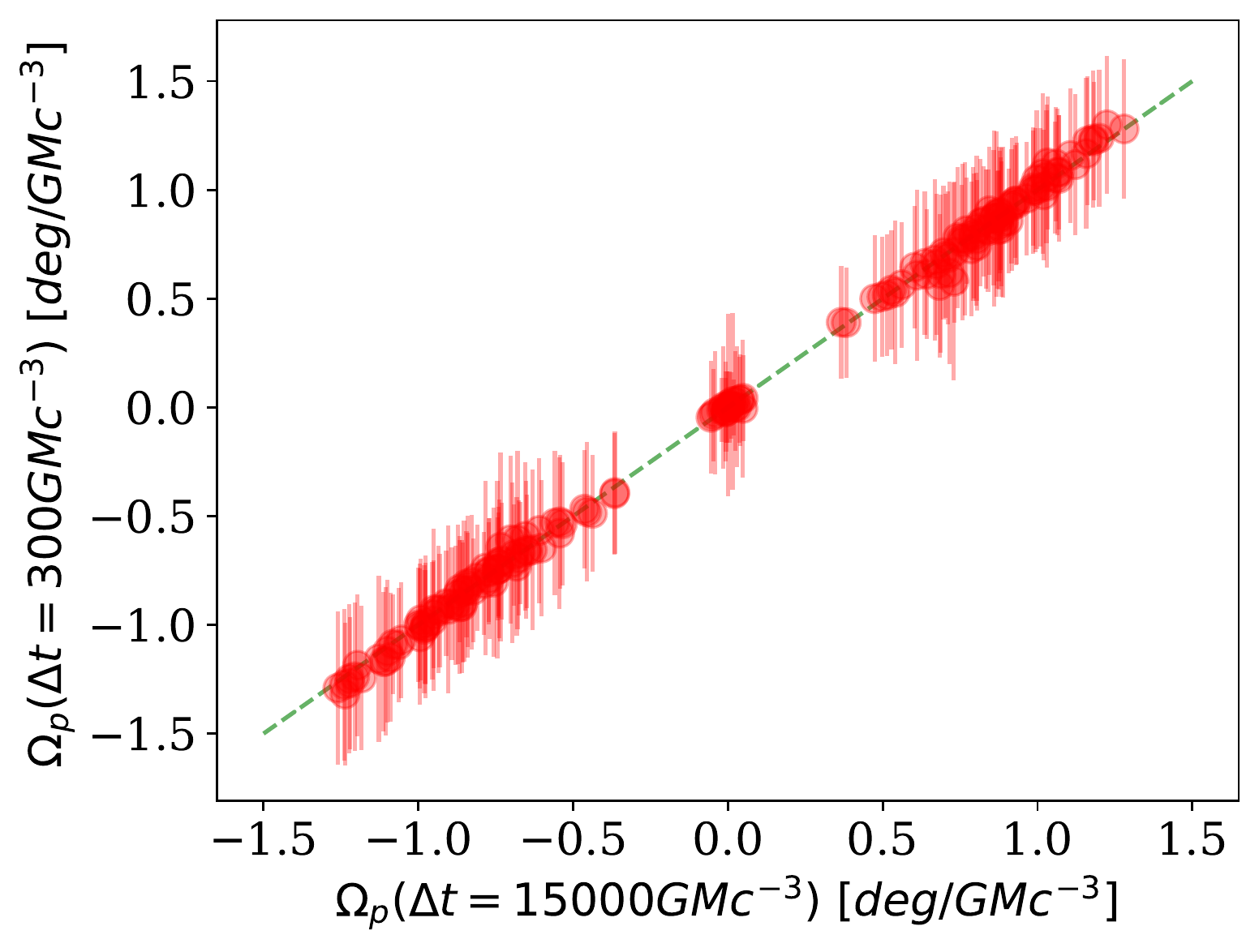}{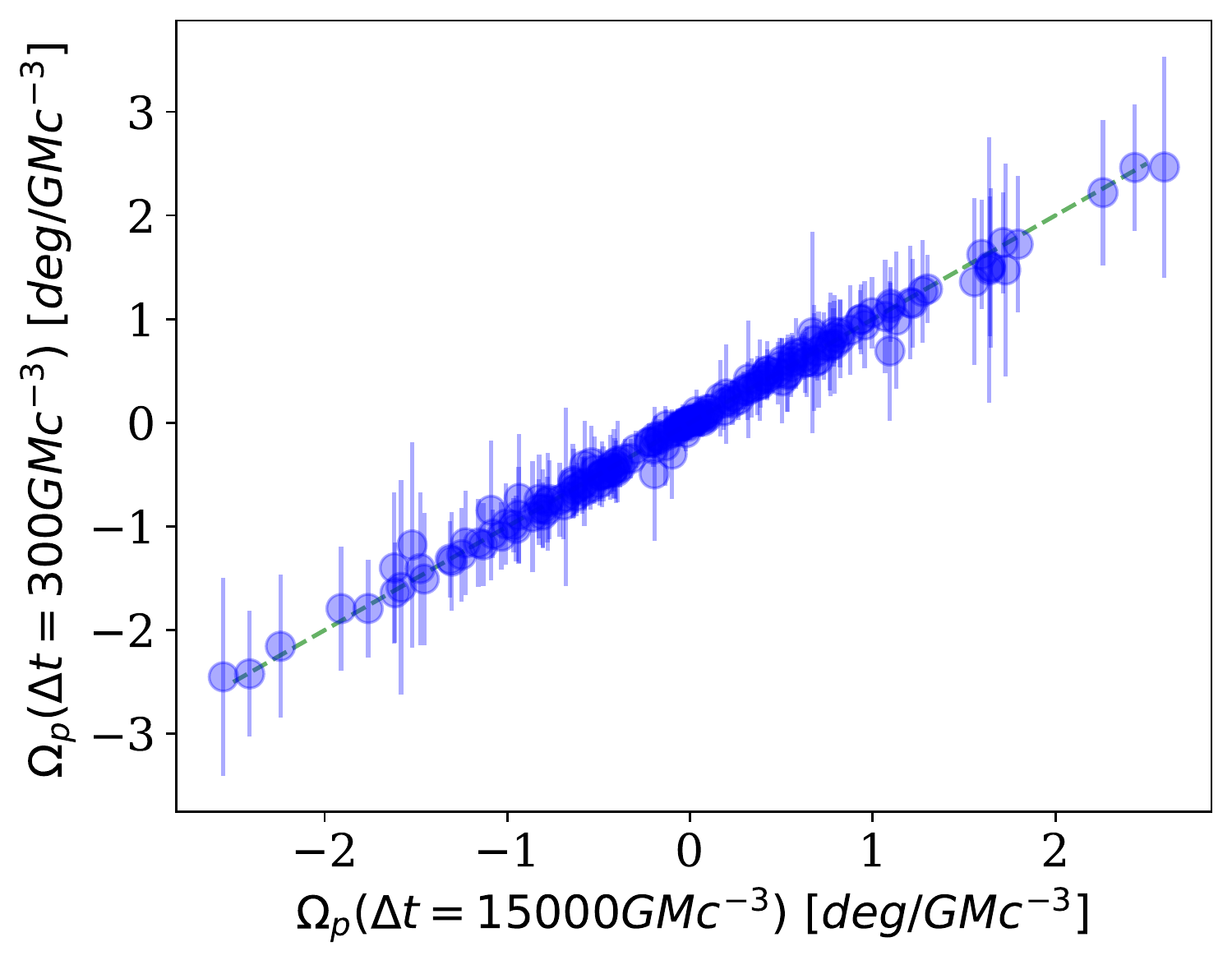}
    \caption{
        $\Omega_p$ found with a longer $10\,GMc^{-3}$ cadence and a shorter $300\,GMc^{-3}$ duration, across all MAD models (left) and SANE models (right). The mean of each set of $300 GMc^{-3}$ subwindows corresponds to the $y$-value, while the error bars denote the standard deviation. If there were no variability in $\Omega_p [300\,GMc^{-3}]$ from that found in the full $15,000\,GMc^{-3}$ window, all points would fall along the dotted green line.      
        \label{fig:observation-like}
    }
\end{figure*}

Observing runs are likely to have shorter duration and lower cadence than our GRMHD movies.  We examined the short-timescale variability of $\Omega_p$ across the model library when measured at $10\, GMc^{-3}$ cadence and $300\,GMc^{-3}$ duration by subdividing the full $1.5 \times 10^4\,GMc^{-3}$ span of each model into subwindows.  The results are shown in Figure \ref{fig:observation-like}, which shows the mean and standard deviation of $\Omega_p$ over subwindows. The average standard deviation is $0.32^\circ$ per $GMc^{-3}$ and the root mean squared variation is $0.06^\circ$ per $GMc^{-3}$.

\section{Conclusion}
\label{sec:conclusion}

This paper proposes a method to measure the rotation of brightness fluctuations in synthetic movies of EHT sources.  We start by measuring surface brightness near the photon ring as a function of $\PA$ and time to produce a so-called cylinder plot.  Rotation manifests in the cylinder plot as features that brighten, change $\PA$, and then fade away.  After normalizing the cylinder plot, we calculate its autocorrelation and use second moments of the autocorrelation to measure the apparent pattern speed $\Omega_p$.

We ran this procedure over the entire Illinois Sgr A* image library, which covers a broad range of plausible configurations for the source plasma, and in every case found the near-horizon pattern speed to be strongly sub-Keplerian, with a mean magnitude of $|\overline{{\Omega}_p}| \approx 1 \,\, \mathrm{deg/GMc^{-3}}$, which is only approximately one-seventh of the expected Keplerian orbital velocity.  This phenomenon can plausibly be attributed to the azimuthal phase velocity of ingoing spiral shocks excited outside the region that produces the bulk of the emission.  Low pattern speeds are a fundamental prediction of GRMHD models.

We also found that $\Omega_p$ depends strongly on inclination, with $\Omega_p$ changing sign as the inclination crosses $90^\circ$.  The pattern speed scales with mass and is only weakly dependent on the spin. The expected pattern speeds are summarized by the fits in Equations~(\ref{eq:fitSANE}) and (\ref{eq:fitMAD}).  

In M87*, a pattern speed measurement would constrain the black hole mass, independent of distance, assuming the accretion flow inclination matches that of the large-scale jet.  Since the sign of the pattern speed follows the accretion flow and the asymmetry of the observed ring follows the black hole spin \citep{M87PaperV}, a measurement of the pattern speed in M87* can distinguish between prograde and retrograde accretion. 

In Sgr A*, where the black hole mass is known to high accuracy from stellar orbit measurements but the accretion flow inclination is not known, a pattern speed measurement would constrain the inclination. \citet{Wielgus_2022_hotspots} measured the linear polarization of millimeter emission immediately following an X-ray flare and found an evolution consistent with clockwise motion on the sky (see also \citealt{Vos_2022}). The GRAVITY Collaboration has measured astrometric motion during infrared flares of Sgr A*, and this motion is also consistent with clockwise rotation on the sky and an approximately face-on inclination \citep{GRAVITY_2018, GRAVITY_2020_flares}. Both of these measurements lead us to expect that $\Omega_p < 0$ in Sgr A*. Notice, however, that GRAVITY measured a super-Keplerian rate of rotation, corresponding to a pattern speed much faster than what we find in this work. 

The accuracy of pattern speed measurements is limited by angular resolution, movie frame rate, movie duration, and the fast light approximation.  We found that a broad range of sampling cadences around $5\,GMc^{-3}$ produces similar pattern speeds.  Subdividing our synthetic movies into shorter-duration movies produces similar but slightly varying pattern speeds.  In a single test model computed with both the fast light approximation and slow light (no approximation), the pattern speeds differ by 19\%.  Finally, the analysis in this paper uses a set of models with similar initial and boundary conditions.  It would be interesting to measure the pattern speed in alternative models (e.g. \citealt{Ressler_2020}, \citealt{White_2020}) since the pattern speed may be uniquely sensitive to  conditions at radii outside the emission region.  

The results presented here suggest that EHT will be able to measure pattern speeds, and that this measurement will provide valuable parameter constraints for M87* and Sgr A*.  Techniques will need to be developed, however, that work directly with data in the $(u,v)$ domain, and which are robust to gaps in $(u,v)$ coverage and irregularly spaced data.  We leave the problem of optimally extracting pattern speeds from EHT data for future work.  

\section{Acknowledgements} \label{sec:Acknowledgements}
This work was supported by NSF grants AST 17-16327 (horizon), OISE 17-43747, and AST 20-34306.  This research used resources of the Oak Ridge Leadership Computing Facility at the Oak Ridge National Laboratory, which is supported by the Office of Science of the U.S. Department of Energy under Contract No. DE-AC05-00OR22725.  This research used resources of the Argonne Leadership Computing Facility, which is a DOE Office of Science User Facility supported under Contract DE-AC02-06CH11357.  This research was done using services provided by the OSG Consortium, which is supported by the National Science Foundation awards \#2030508 and \#1836650.  This research is part of the Delta research computing project, which is supported by the National Science Foundation (award OCI 2005572), and the State of Illinois. Delta is a joint effort of the University of Illinois at Urbana-Champaign and its National Center for Supercomputing Applications.

This work was also supported in part by Perimeter Institute for Theoretical Physics.  Research at Perimeter Institute is supported by the Government of Canada through the Department of Innovation, Science and Economic Development Canada, and by the Province of Ontario through the Ministry of Economic Development, Job Creation and Trade.  A.E.B. thanks the Delaney Family for their generous financial support via the Delaney Family John A. Wheeler Chair at Perimeter Institute.  A.E.B. receives additional financial support from the Natural Sciences and Engineering Research Council of Canada through a Discovery Grant.

We are grateful to George Wong for providing the slow light models used in Section \ref{subsec:slowlight}. We thank Steve Balbus, Alex Lupsasca, Maciek Wielgus, George Wong, and the anonymous referee for comments that greatly improved this paper.

\vspace{5mm}

\software{KHARMA, \texttt{ipole}}

\appendix
\section{Pattern Speed Fits From Sgr A* Library}
\label{subsec:appendix}

The main text summarized $\Omega_p$ measurements in the Sgr A* library using fits (see Equations~\ref{eq:fitMAD} and \ref{eq:fitSANE}).  This appendix provides $\Omega_p$ and $\Omega_{p,\mathrm{FIT}}$ for each model in the Illinois component of the Sgr A* model library.  The units for $\Omega_p$ and $\Omega_{p,\mathrm{FIT}}$ are degrees per $GMc^{-3}$.

The model library surveys across four parameters: spin ($\spin = $-0.94, -0.5, 0, 0.5, and 0.94), magnetization (MAD or SANE), inclination ($i =$ 10, 30, 50, 70, 90, 110, 130, 150, and 170$^\circ$), and electron distribution parameter $\rhigh$ ($\rhigh = $1, 10, 40, and 160). 

\startlongtable
\begin{deluxetable}{cccccc}
\tablecaption{Pattern Speeds from the Sgr A* Library} 
\label{tab:long} 
\tablehead{\colhead{MAD$/$SANE} & \colhead{$\spin$} & \colhead{$i$ [deg]} & \colhead{$\rhigh$} & \colhead{$\Omega_p$} & \colhead{$\Omega_{p,\mathrm{FIT}}$} }  
\startdata
MAD & -0.94 & 10.0 & 1.0 & 0.91 & 1.04 \\
MAD & -0.94 & 30.0 & 1.0 & 0.87 & 0.92 \\
MAD & -0.94 & 50.0 & 1.0 & 0.8 & 0.68 \\
MAD & -0.94 & 70.0 & 1.0 & 0.64 & 0.36 \\
MAD & -0.94 & 90.0 & 1.0 & -0.01 & 0.0 \\
MAD & -0.94 & 110.0 & 1.0 & -0.61 & -0.36 \\
MAD & -0.94 & 130.0 & 1.0 & -0.74 & -0.68 \\
MAD & -0.94 & 150.0 & 1.0 & -0.81 & -0.92 \\
MAD & -0.94 & 170.0 & 1.0 & -0.87 & -1.04 \\
MAD & -0.94 & 10.0 & 10.0 & 0.87 & 0.99 \\
MAD & -0.94 & 30.0 & 10.0 & 0.84 & 0.87 \\
MAD & -0.94 & 50.0 & 10.0 & 0.79 & 0.65 \\
MAD & -0.94 & 70.0 & 10.0 & 0.56 & 0.34 \\
MAD & -0.94 & 90.0 & 10.0 & -0.04 & 0.0 \\
MAD & -0.94 & 110.0 & 10.0 & -0.56 & -0.34 \\
MAD & -0.94 & 130.0 & 10.0 & -0.73 & -0.65 \\
MAD & -0.94 & 150.0 & 10.0 & -0.77 & -0.87 \\
MAD & -0.94 & 170.0 & 10.0 & -0.84 & -0.99 \\
MAD & -0.94 & 10.0 & 40.0 & 0.79 & 0.96 \\
MAD & -0.94 & 30.0 & 40.0 & 0.8 & 0.84 \\
MAD & -0.94 & 50.0 & 40.0 & 0.7 & 0.63 \\
MAD & -0.94 & 70.0 & 40.0 & 0.53 & 0.33 \\
MAD & -0.94 & 90.0 & 40.0 & -0.05 & 0.0 \\
MAD & -0.94 & 110.0 & 40.0 & -0.54 & -0.33 \\
MAD & -0.94 & 130.0 & 40.0 & -0.68 & -0.63 \\
MAD & -0.94 & 150.0 & 40.0 & -0.74 & -0.84 \\
MAD & -0.94 & 170.0 & 40.0 & -0.77 & -0.96 \\
MAD & -0.94 & 10.0 & 160.0 & 0.7 & 0.93 \\
MAD & -0.94 & 30.0 & 160.0 & 0.71 & 0.81 \\
MAD & -0.94 & 50.0 & 160.0 & 0.67 & 0.6 \\
MAD & -0.94 & 70.0 & 160.0 & 0.5 & 0.32 \\
MAD & -0.94 & 90.0 & 160.0 & -0.05 & 0.0 \\
MAD & -0.94 & 110.0 & 160.0 & -0.53 & -0.32 \\
MAD & -0.94 & 130.0 & 160.0 & -0.65 & -0.6 \\
MAD & -0.94 & 150.0 & 160.0 & -0.67 & -0.81 \\
MAD & -0.94 & 170.0 & 160.0 & -0.69 & -0.93 \\
MAD & -0.5 & 10.0 & 1.0 & 0.99 & 1.11 \\
MAD & -0.5 & 30.0 & 1.0 & 0.93 & 0.98 \\
MAD & -0.5 & 50.0 & 1.0 & 0.86 & 0.73 \\
MAD & -0.5 & 70.0 & 1.0 & 0.76 & 0.39 \\
MAD & -0.5 & 90.0 & 1.0 & 0.01 & 0.0 \\
MAD & -0.5 & 110.0 & 1.0 & -0.75 & -0.39 \\
MAD & -0.5 & 130.0 & 1.0 & -0.87 & -0.73 \\
MAD & -0.5 & 150.0 & 1.0 & -0.94 & -0.98 \\
MAD & -0.5 & 170.0 & 1.0 & -1.0 & -1.11 \\
MAD & -0.5 & 10.0 & 10.0 & 0.99 & 1.06 \\
MAD & -0.5 & 30.0 & 10.0 & 0.92 & 0.93 \\
MAD & -0.5 & 50.0 & 10.0 & 0.88 & 0.69 \\
MAD & -0.5 & 70.0 & 10.0 & 0.77 & 0.37 \\
MAD & -0.5 & 90.0 & 10.0 & -0.0 & 0.0 \\
MAD & -0.5 & 110.0 & 10.0 & -0.78 & -0.37 \\
MAD & -0.5 & 130.0 & 10.0 & -0.88 & -0.69 \\
MAD & -0.5 & 150.0 & 10.0 & -0.95 & -0.93 \\
MAD & -0.5 & 170.0 & 10.0 & -1.0 & -1.06 \\
MAD & -0.5 & 10.0 & 40.0 & 0.93 & 1.03 \\
MAD & -0.5 & 30.0 & 40.0 & 0.88 & 0.9 \\
MAD & -0.5 & 50.0 & 40.0 & 0.82 & 0.67 \\
MAD & -0.5 & 70.0 & 40.0 & 0.73 & 0.36 \\
MAD & -0.5 & 90.0 & 40.0 & 0.05 & 0.0 \\
MAD & -0.5 & 110.0 & 40.0 & -0.7 & -0.36 \\
MAD & -0.5 & 130.0 & 40.0 & -0.84 & -0.67 \\
MAD & -0.5 & 150.0 & 40.0 & -0.91 & -0.9 \\
MAD & -0.5 & 170.0 & 40.0 & -0.93 & -1.03 \\
MAD & -0.5 & 10.0 & 160.0 & 0.83 & 0.99 \\
MAD & -0.5 & 30.0 & 160.0 & 0.77 & 0.87 \\
MAD & -0.5 & 50.0 & 160.0 & 0.76 & 0.65 \\
MAD & -0.5 & 70.0 & 160.0 & 0.6 & 0.34 \\
MAD & -0.5 & 90.0 & 160.0 & 0.03 & 0.0 \\
MAD & -0.5 & 110.0 & 160.0 & -0.65 & -0.34 \\
MAD & -0.5 & 130.0 & 160.0 & -0.74 & -0.65 \\
MAD & -0.5 & 150.0 & 160.0 & -0.79 & -0.87 \\
MAD & -0.5 & 170.0 & 160.0 & -0.82 & -0.99 \\
MAD & 0.0 & 10.0 & 1.0 & 1.07 & 1.19 \\
MAD & 0.0 & 30.0 & 1.0 & 1.02 & 1.04 \\
MAD & 0.0 & 50.0 & 1.0 & 0.89 & 0.78 \\
MAD & 0.0 & 70.0 & 1.0 & 0.71 & 0.41 \\
MAD & 0.0 & 90.0 & 1.0 & 0.02 & 0.0 \\
MAD & 0.0 & 110.0 & 1.0 & -0.68 & -0.41 \\
MAD & 0.0 & 130.0 & 1.0 & -0.84 & -0.78 \\
MAD & 0.0 & 150.0 & 1.0 & -0.99 & -1.04 \\
MAD & 0.0 & 170.0 & 1.0 & -1.05 & -1.19 \\
MAD & 0.0 & 10.0 & 10.0 & 1.07 & 1.13 \\
MAD & 0.0 & 30.0 & 10.0 & 1.01 & 1.0 \\
MAD & 0.0 & 50.0 & 10.0 & 0.88 & 0.74 \\
MAD & 0.0 & 70.0 & 10.0 & 0.73 & 0.39 \\
MAD & 0.0 & 90.0 & 10.0 & 0.04 & 0.0 \\
MAD & 0.0 & 110.0 & 10.0 & -0.73 & -0.39 \\
MAD & 0.0 & 130.0 & 10.0 & -0.85 & -0.74 \\
MAD & 0.0 & 150.0 & 10.0 & -0.98 & -1.0 \\
MAD & 0.0 & 170.0 & 10.0 & -1.08 & -1.13 \\
MAD & 0.0 & 10.0 & 40.0 & 1.06 & 1.1 \\
MAD & 0.0 & 30.0 & 40.0 & 0.99 & 0.97 \\
MAD & 0.0 & 50.0 & 40.0 & 0.88 & 0.72 \\
MAD & 0.0 & 70.0 & 40.0 & 0.69 & 0.38 \\
MAD & 0.0 & 90.0 & 40.0 & 0.04 & 0.0 \\
MAD & 0.0 & 110.0 & 40.0 & -0.7 & -0.38 \\
MAD & 0.0 & 130.0 & 40.0 & -0.83 & -0.72 \\
MAD & 0.0 & 150.0 & 40.0 & -0.95 & -0.97 \\
MAD & 0.0 & 170.0 & 40.0 & -1.06 & -1.1 \\
MAD & 0.0 & 10.0 & 160.0 & 0.97 & 1.07 \\
MAD & 0.0 & 30.0 & 160.0 & 0.92 & 0.94 \\
MAD & 0.0 & 50.0 & 160.0 & 0.83 & 0.7 \\
MAD & 0.0 & 70.0 & 160.0 & 0.68 & 0.37 \\
MAD & 0.0 & 90.0 & 160.0 & 0.05 & 0.0 \\
MAD & 0.0 & 110.0 & 160.0 & -0.65 & -0.37 \\
MAD & 0.0 & 130.0 & 160.0 & -0.77 & -0.7 \\
MAD & 0.0 & 150.0 & 160.0 & -0.89 & -0.94 \\
MAD & 0.0 & 170.0 & 160.0 & -0.98 & -1.07 \\
MAD & 0.5 & 10.0 & 1.0 & 1.12 & 1.26 \\
MAD & 0.5 & 30.0 & 1.0 & 1.03 & 1.11 \\
MAD & 0.5 & 50.0 & 1.0 & 0.87 & 0.82 \\
MAD & 0.5 & 70.0 & 1.0 & 0.67 & 0.44 \\
MAD & 0.5 & 90.0 & 1.0 & -0.01 & 0.0 \\
MAD & 0.5 & 110.0 & 1.0 & -0.63 & -0.44 \\
MAD & 0.5 & 130.0 & 1.0 & -0.95 & -0.82 \\
MAD & 0.5 & 150.0 & 1.0 & -1.1 & -1.11 \\
MAD & 0.5 & 170.0 & 1.0 & -1.2 & -1.26 \\
MAD & 0.5 & 10.0 & 10.0 & 1.16 & 1.21 \\
MAD & 0.5 & 30.0 & 10.0 & 1.02 & 1.06 \\
MAD & 0.5 & 50.0 & 10.0 & 0.86 & 0.79 \\
MAD & 0.5 & 70.0 & 10.0 & 0.61 & 0.42 \\
MAD & 0.5 & 90.0 & 10.0 & -0.0 & 0.0 \\
MAD & 0.5 & 110.0 & 10.0 & -0.54 & -0.42 \\
MAD & 0.5 & 130.0 & 10.0 & -0.9 & -0.79 \\
MAD & 0.5 & 150.0 & 10.0 & -1.13 & -1.06 \\
MAD & 0.5 & 170.0 & 10.0 & -1.22 & -1.21 \\
MAD & 0.5 & 10.0 & 40.0 & 1.16 & 1.18 \\
MAD & 0.5 & 30.0 & 40.0 & 1.03 & 1.04 \\
MAD & 0.5 & 50.0 & 40.0 & 0.84 & 0.77 \\
MAD & 0.5 & 70.0 & 40.0 & 0.54 & 0.41 \\
MAD & 0.5 & 90.0 & 40.0 & 0.01 & 0.0 \\
MAD & 0.5 & 110.0 & 40.0 & -0.46 & -0.41 \\
MAD & 0.5 & 130.0 & 40.0 & -0.86 & -0.77 \\
MAD & 0.5 & 150.0 & 40.0 & -1.11 & -1.04 \\
MAD & 0.5 & 170.0 & 40.0 & -1.24 & -1.18 \\
MAD & 0.5 & 10.0 & 160.0 & 1.18 & 1.14 \\
MAD & 0.5 & 30.0 & 160.0 & 1.06 & 1.01 \\
MAD & 0.5 & 50.0 & 160.0 & 0.88 & 0.75 \\
MAD & 0.5 & 70.0 & 160.0 & 0.51 & 0.4 \\
MAD & 0.5 & 90.0 & 160.0 & 0.01 & 0.0 \\
MAD & 0.5 & 110.0 & 160.0 & -0.44 & -0.4 \\
MAD & 0.5 & 130.0 & 160.0 & -0.86 & -0.75 \\
MAD & 0.5 & 150.0 & 160.0 & -1.11 & -1.01 \\
MAD & 0.5 & 170.0 & 160.0 & -1.22 & -1.14 \\
MAD & 0.94 & 10.0 & 1.0 & 1.22 & 1.33 \\
MAD & 0.94 & 30.0 & 1.0 & 1.11 & 1.17 \\
MAD & 0.94 & 50.0 & 1.0 & 0.93 & 0.87 \\
MAD & 0.94 & 70.0 & 1.0 & 0.64 & 0.46 \\
MAD & 0.94 & 90.0 & 1.0 & -0.01 & 0.0 \\
MAD & 0.94 & 110.0 & 1.0 & -0.6 & -0.46 \\
MAD & 0.94 & 130.0 & 1.0 & -0.87 & -0.87 \\
MAD & 0.94 & 150.0 & 1.0 & -1.09 & -1.17 \\
MAD & 0.94 & 170.0 & 1.0 & -1.24 & -1.33 \\
MAD & 0.94 & 10.0 & 10.0 & 1.18 & 1.28 \\
MAD & 0.94 & 30.0 & 10.0 & 1.03 & 1.12 \\
MAD & 0.94 & 50.0 & 10.0 & 0.8 & 0.83 \\
MAD & 0.94 & 70.0 & 10.0 & 0.47 & 0.44 \\
MAD & 0.94 & 90.0 & 10.0 & -0.01 & 0.0 \\
MAD & 0.94 & 110.0 & 10.0 & -0.46 & -0.44 \\
MAD & 0.94 & 130.0 & 10.0 & -0.76 & -0.83 \\
MAD & 0.94 & 150.0 & 10.0 & -0.99 & -1.12 \\
MAD & 0.94 & 170.0 & 10.0 & -1.18 & -1.28 \\
MAD & 0.94 & 10.0 & 40.0 & 1.2 & 1.24 \\
MAD & 0.94 & 30.0 & 40.0 & 1.0 & 1.09 \\
MAD & 0.94 & 50.0 & 40.0 & 0.74 & 0.81 \\
MAD & 0.94 & 70.0 & 40.0 & 0.37 & 0.43 \\
MAD & 0.94 & 90.0 & 40.0 & -0.02 & 0.0 \\
MAD & 0.94 & 110.0 & 40.0 & -0.37 & -0.43 \\
MAD & 0.94 & 130.0 & 40.0 & -0.68 & -0.81 \\
MAD & 0.94 & 150.0 & 40.0 & -0.98 & -1.09 \\
MAD & 0.94 & 170.0 & 40.0 & -1.21 & -1.24 \\
MAD & 0.94 & 10.0 & 160.0 & 1.28 & 1.21 \\
MAD & 0.94 & 30.0 & 160.0 & 1.06 & 1.07 \\
MAD & 0.94 & 50.0 & 160.0 & 0.79 & 0.79 \\
MAD & 0.94 & 70.0 & 160.0 & 0.38 & 0.42 \\
MAD & 0.94 & 90.0 & 160.0 & 0.02 & 0.0 \\
MAD & 0.94 & 110.0 & 160.0 & -0.37 & -0.42 \\
MAD & 0.94 & 130.0 & 160.0 & -0.74 & -0.79 \\
MAD & 0.94 & 150.0 & 160.0 & -0.98 & -1.07 \\
MAD & 0.94 & 170.0 & 160.0 & -1.26 & -1.21 \\
SANE & -0.94 & 10.0 & 1.0 & 0.62 & 1.15 \\
SANE & -0.94 & 30.0 & 1.0 & 0.56 & 1.01 \\
SANE & -0.94 & 50.0 & 1.0 & 0.5 & 0.75 \\
SANE & -0.94 & 70.0 & 1.0 & 0.42 & 0.4 \\
SANE & -0.94 & 90.0 & 1.0 & -0.01 & 0.0 \\
SANE & -0.94 & 110.0 & 1.0 & -0.43 & -0.4 \\
SANE & -0.94 & 130.0 & 1.0 & -0.5 & -0.75 \\
SANE & -0.94 & 150.0 & 1.0 & -0.54 & -1.01 \\
SANE & -0.94 & 170.0 & 1.0 & -0.62 & -1.15 \\
SANE & -0.94 & 10.0 & 10.0 & 1.1 & 0.79 \\
SANE & -0.94 & 30.0 & 10.0 & 0.93 & 0.69 \\
SANE & -0.94 & 50.0 & 10.0 & 0.78 & 0.51 \\
SANE & -0.94 & 70.0 & 10.0 & 0.63 & 0.27 \\
SANE & -0.94 & 90.0 & 10.0 & 0.05 & 0.0 \\
SANE & -0.94 & 110.0 & 10.0 & -0.64 & -0.27 \\
SANE & -0.94 & 130.0 & 10.0 & -0.83 & -0.51 \\
SANE & -0.94 & 150.0 & 10.0 & -0.97 & -0.69 \\
SANE & -0.94 & 170.0 & 10.0 & -1.08 & -0.79 \\
SANE & -0.94 & 10.0 & 40.0 & 0.7 & 0.57 \\
SANE & -0.94 & 30.0 & 40.0 & 0.62 & 0.5 \\
SANE & -0.94 & 50.0 & 40.0 & 0.43 & 0.37 \\
SANE & -0.94 & 70.0 & 40.0 & 0.21 & 0.2 \\
SANE & -0.94 & 90.0 & 40.0 & 0.01 & 0.0 \\
SANE & -0.94 & 110.0 & 40.0 & -0.22 & -0.2 \\
SANE & -0.94 & 130.0 & 40.0 & -0.49 & -0.37 \\
SANE & -0.94 & 150.0 & 40.0 & -0.64 & -0.5 \\
SANE & -0.94 & 170.0 & 40.0 & -0.62 & -0.57 \\
SANE & -0.94 & 10.0 & 160.0 & 0.42 & 0.35 \\
SANE & -0.94 & 30.0 & 160.0 & 0.4 & 0.3 \\
SANE & -0.94 & 50.0 & 160.0 & 0.32 & 0.23 \\
SANE & -0.94 & 70.0 & 160.0 & 0.11 & 0.12 \\
SANE & -0.94 & 90.0 & 160.0 & 0.01 & 0.0 \\
SANE & -0.94 & 110.0 & 160.0 & -0.1 & -0.12 \\
SANE & -0.94 & 130.0 & 160.0 & -0.38 & -0.23 \\
SANE & -0.94 & 150.0 & 160.0 & -0.45 & -0.3 \\
SANE & -0.94 & 170.0 & 160.0 & -0.41 & -0.35 \\
SANE & -0.5 & 10.0 & 1.0 & 0.79 & 1.31 \\
SANE & -0.5 & 30.0 & 1.0 & 0.68 & 1.15 \\
SANE & -0.5 & 50.0 & 1.0 & 0.58 & 0.85 \\
SANE & -0.5 & 70.0 & 1.0 & 0.42 & 0.45 \\
SANE & -0.5 & 90.0 & 1.0 & 0.02 & 0.0 \\
SANE & -0.5 & 110.0 & 1.0 & -0.4 & -0.45 \\
SANE & -0.5 & 130.0 & 1.0 & -0.58 & -0.85 \\
SANE & -0.5 & 150.0 & 1.0 & -0.69 & -1.15 \\
SANE & -0.5 & 170.0 & 1.0 & -0.81 & -1.31 \\
SANE & -0.5 & 10.0 & 10.0 & 1.3 & 0.94 \\
SANE & -0.5 & 30.0 & 10.0 & 1.1 & 0.83 \\
SANE & -0.5 & 50.0 & 10.0 & 0.82 & 0.61 \\
SANE & -0.5 & 70.0 & 10.0 & 0.54 & 0.33 \\
SANE & -0.5 & 90.0 & 10.0 & 0.1 & 0.0 \\
SANE & -0.5 & 110.0 & 10.0 & -0.52 & -0.33 \\
SANE & -0.5 & 130.0 & 10.0 & -0.81 & -0.61 \\
SANE & -0.5 & 150.0 & 10.0 & -1.13 & -0.83 \\
SANE & -0.5 & 170.0 & 10.0 & -1.31 & -0.94 \\
SANE & -0.5 & 10.0 & 40.0 & 0.96 & 0.72 \\
SANE & -0.5 & 30.0 & 40.0 & 0.73 & 0.63 \\
SANE & -0.5 & 50.0 & 40.0 & 0.39 & 0.47 \\
SANE & -0.5 & 70.0 & 40.0 & 0.19 & 0.25 \\
SANE & -0.5 & 90.0 & 40.0 & 0.03 & 0.0 \\
SANE & -0.5 & 110.0 & 40.0 & -0.17 & -0.25 \\
SANE & -0.5 & 130.0 & 40.0 & -0.35 & -0.47 \\
SANE & -0.5 & 150.0 & 40.0 & -0.77 & -0.63 \\
SANE & -0.5 & 170.0 & 40.0 & -1.01 & -0.72 \\
SANE & -0.5 & 10.0 & 160.0 & 0.56 & 0.5 \\
SANE & -0.5 & 30.0 & 160.0 & 0.55 & 0.44 \\
SANE & -0.5 & 50.0 & 160.0 & 0.29 & 0.33 \\
SANE & -0.5 & 70.0 & 160.0 & 0.1 & 0.17 \\
SANE & -0.5 & 90.0 & 160.0 & 0.02 & 0.0 \\
SANE & -0.5 & 110.0 & 160.0 & -0.08 & -0.17 \\
SANE & -0.5 & 130.0 & 160.0 & -0.32 & -0.33 \\
SANE & -0.5 & 150.0 & 160.0 & -0.58 & -0.44 \\
SANE & -0.5 & 170.0 & 160.0 & -0.59 & -0.5 \\
SANE & 0.0 & 10.0 & 1.0 & 1.0 & 1.48 \\
SANE & 0.0 & 30.0 & 1.0 & 0.94 & 1.3 \\
SANE & 0.0 & 50.0 & 1.0 & 0.82 & 0.97 \\
SANE & 0.0 & 70.0 & 1.0 & 0.49 & 0.51 \\
SANE & 0.0 & 90.0 & 1.0 & -0.04 & 0.0 \\
SANE & 0.0 & 110.0 & 1.0 & -0.5 & -0.51 \\
SANE & 0.0 & 130.0 & 1.0 & -0.78 & -0.97 \\
SANE & 0.0 & 150.0 & 1.0 & -0.95 & -1.3 \\
SANE & 0.0 & 170.0 & 1.0 & -1.03 & -1.48 \\
SANE & 0.0 & 10.0 & 10.0 & 1.27 & 1.12 \\
SANE & 0.0 & 30.0 & 10.0 & 1.07 & 0.98 \\
SANE & 0.0 & 50.0 & 10.0 & 0.77 & 0.73 \\
SANE & 0.0 & 70.0 & 10.0 & 0.31 & 0.39 \\
SANE & 0.0 & 90.0 & 10.0 & -0.07 & 0.0 \\
SANE & 0.0 & 110.0 & 10.0 & -0.41 & -0.39 \\
SANE & 0.0 & 130.0 & 10.0 & -0.83 & -0.73 \\
SANE & 0.0 & 150.0 & 10.0 & -1.16 & -0.98 \\
SANE & 0.0 & 170.0 & 10.0 & -1.3 & -1.12 \\
SANE & 0.0 & 10.0 & 40.0 & 1.22 & 0.9 \\
SANE & 0.0 & 30.0 & 40.0 & 0.76 & 0.79 \\
SANE & 0.0 & 50.0 & 40.0 & 0.37 & 0.59 \\
SANE & 0.0 & 70.0 & 40.0 & 0.09 & 0.31 \\
SANE & 0.0 & 90.0 & 40.0 & -0.06 & 0.0 \\
SANE & 0.0 & 110.0 & 40.0 & -0.2 & -0.31 \\
SANE & 0.0 & 130.0 & 40.0 & -0.44 & -0.59 \\
SANE & 0.0 & 150.0 & 40.0 & -0.94 & -0.79 \\
SANE & 0.0 & 170.0 & 40.0 & -1.25 & -0.9 \\
SANE & 0.0 & 10.0 & 160.0 & 0.79 & 0.68 \\
SANE & 0.0 & 30.0 & 160.0 & 0.48 & 0.6 \\
SANE & 0.0 & 50.0 & 160.0 & 0.24 & 0.44 \\
SANE & 0.0 & 70.0 & 160.0 & 0.08 & 0.24 \\
SANE & 0.0 & 90.0 & 160.0 & -0.02 & 0.0 \\
SANE & 0.0 & 110.0 & 160.0 & -0.12 & -0.24 \\
SANE & 0.0 & 130.0 & 160.0 & -0.2 & -0.44 \\
SANE & 0.0 & 150.0 & 160.0 & -0.47 & -0.6 \\
SANE & 0.0 & 170.0 & 160.0 & -0.71 & -0.68 \\
SANE & 0.5 & 10.0 & 1.0 & 1.71 & 1.66 \\
SANE & 0.5 & 30.0 & 1.0 & 1.6 & 1.46 \\
SANE & 0.5 & 50.0 & 1.0 & 1.21 & 1.08 \\
SANE & 0.5 & 70.0 & 1.0 & 0.64 & 0.58 \\
SANE & 0.5 & 90.0 & 1.0 & -0.0 & 0.0 \\
SANE & 0.5 & 110.0 & 1.0 & -0.64 & -0.58 \\
SANE & 0.5 & 130.0 & 1.0 & -1.23 & -1.08 \\
SANE & 0.5 & 150.0 & 1.0 & -1.61 & -1.46 \\
SANE & 0.5 & 170.0 & 1.0 & -1.76 & -1.66 \\
SANE & 0.5 & 10.0 & 10.0 & 1.8 & 1.29 \\
SANE & 0.5 & 30.0 & 10.0 & 1.64 & 1.14 \\
SANE & 0.5 & 50.0 & 10.0 & 1.13 & 0.84 \\
SANE & 0.5 & 70.0 & 10.0 & 0.5 & 0.45 \\
SANE & 0.5 & 90.0 & 10.0 & -0.05 & 0.0 \\
SANE & 0.5 & 110.0 & 10.0 & -0.54 & -0.45 \\
SANE & 0.5 & 130.0 & 10.0 & -1.09 & -0.84 \\
SANE & 0.5 & 150.0 & 10.0 & -1.62 & -1.14 \\
SANE & 0.5 & 170.0 & 10.0 & -1.91 & -1.29 \\
SANE & 0.5 & 10.0 & 40.0 & 1.64 & 1.07 \\
SANE & 0.5 & 30.0 & 40.0 & 0.7 & 0.94 \\
SANE & 0.5 & 50.0 & 40.0 & 0.34 & 0.7 \\
SANE & 0.5 & 70.0 & 40.0 & 0.23 & 0.37 \\
SANE & 0.5 & 90.0 & 40.0 & -0.04 & 0.0 \\
SANE & 0.5 & 110.0 & 40.0 & -0.23 & -0.37 \\
SANE & 0.5 & 130.0 & 40.0 & -0.42 & -0.7 \\
SANE & 0.5 & 150.0 & 40.0 & -0.78 & -0.94 \\
SANE & 0.5 & 170.0 & 40.0 & -1.45 & -1.07 \\
SANE & 0.5 & 10.0 & 160.0 & 0.88 & 0.85 \\
SANE & 0.5 & 30.0 & 160.0 & 0.53 & 0.75 \\
SANE & 0.5 & 50.0 & 160.0 & 0.27 & 0.56 \\
SANE & 0.5 & 70.0 & 160.0 & 0.08 & 0.3 \\
SANE & 0.5 & 90.0 & 160.0 & -0.03 & 0.0 \\
SANE & 0.5 & 110.0 & 160.0 & -0.17 & -0.3 \\
SANE & 0.5 & 130.0 & 160.0 & -0.3 & -0.56 \\
SANE & 0.5 & 150.0 & 160.0 & -0.48 & -0.75 \\
SANE & 0.5 & 170.0 & 160.0 & -0.86 & -0.85 \\
SANE & 0.94 & 10.0 & 1.0 & 2.43 & 1.81 \\
SANE & 0.94 & 30.0 & 1.0 & 2.26 & 1.6 \\
SANE & 0.94 & 50.0 & 1.0 & 1.56 & 1.18 \\
SANE & 0.94 & 70.0 & 1.0 & 0.68 & 0.63 \\
SANE & 0.94 & 90.0 & 1.0 & 0.03 & 0.0 \\
SANE & 0.94 & 110.0 & 1.0 & -0.58 & -0.63 \\
SANE & 0.94 & 130.0 & 1.0 & -1.48 & -1.18 \\
SANE & 0.94 & 150.0 & 1.0 & -2.24 & -1.6 \\
SANE & 0.94 & 170.0 & 1.0 & -2.41 & -1.81 \\
SANE & 0.94 & 10.0 & 10.0 & 2.6 & 1.45 \\
SANE & 0.94 & 30.0 & 10.0 & 1.64 & 1.27 \\
SANE & 0.94 & 50.0 & 10.0 & 0.67 & 0.95 \\
SANE & 0.94 & 70.0 & 10.0 & 0.2 & 0.5 \\
SANE & 0.94 & 90.0 & 10.0 & -0.02 & 0.0 \\
SANE & 0.94 & 110.0 & 10.0 & -0.14 & -0.5 \\
SANE & 0.94 & 130.0 & 10.0 & -0.68 & -0.95 \\
SANE & 0.94 & 150.0 & 10.0 & -1.58 & -1.27 \\
SANE & 0.94 & 170.0 & 10.0 & -2.55 & -1.45 \\
SANE & 0.94 & 10.0 & 40.0 & 1.73 & 1.23 \\
SANE & 0.94 & 30.0 & 40.0 & 0.32 & 1.08 \\
SANE & 0.94 & 50.0 & 40.0 & 0.16 & 0.8 \\
SANE & 0.94 & 70.0 & 40.0 & 0.04 & 0.43 \\
SANE & 0.94 & 90.0 & 40.0 & -0.01 & 0.0 \\
SANE & 0.94 & 110.0 & 40.0 & -0.02 & -0.43 \\
SANE & 0.94 & 130.0 & 40.0 & -0.1 & -0.8 \\
SANE & 0.94 & 150.0 & 40.0 & -0.2 & -1.08 \\ 
SANE & 0.94 & 170.0 & 40.0 & -1.52 & -1.23 \\ 
SANE & 0.94 & 10.0 & 160.0 & 1.09 & 1.01 \\ 
SANE & 0.94 & 30.0 & 160.0 & 0.38 & 0.89 \\ 
SANE & 0.94 & 50.0 & 160.0 & 0.18 & 0.66 \\ 
SANE & 0.94 & 70.0 & 160.0 & 0.07 & 0.35 \\ 
SANE & 0.94 & 90.0 & 160.0 & 0.06 & 0.0 \\ 
SANE & 0.94 & 110.0 & 160.0 & -0.13 & -0.35 \\ 
SANE & 0.94 & 130.0 & 160.0 & -0.19 & -0.66 \\ 
SANE & 0.94 & 150.0 & 160.0 & -0.39 & -0.89 \\ 
SANE & 0.94 & 170.0 & 160.0 & -0.93 & -1.01
\enddata
\end{deluxetable}

\clearpage
\bibliography{main}{}

\begin{thebibliography}{}
\expandafter\ifx\csname natexlab\endcsname\relax\def\natexlab#1{#1}\fi
\providecommand{\url}[1]{\href{#1}{#1}}
\providecommand{\dodoi}[1]{doi:~\href{http://doi.org/#1}{\nolinkurl{#1}}}
\providecommand{\doeprint}[1]{\href{http://ascl.net/#1}{\nolinkurl{http://ascl.net/#1}}}
\providecommand{\doarXiv}[1]{\href{https://arxiv.org/abs/#1}{\nolinkurl{https://arxiv.org/abs/#1}}}

\bibitem[{Broderick \& Loeb(2006)}]{Broderick_2006}
Broderick, A.~E., \& Loeb, A. 2006, Monthly Notices of the Royal Astronomical
  Society, 367, 905, \dodoi{10.1111/j.1365-2966.2006.10152.x}

\bibitem[{Conroy {et~al.}(2023)Conroy, Baubock, \& Gammie}]{cylinder_clean}
Conroy, N., Baubock, M., \& Gammie, C. 2023, Cylinder\_Clean.py, v1.0,  Zenodo,
  \dodoi{10.5281/zenodo.7809121}

\bibitem[{Do {et~al.}(2019)Do, Hees, Ghez, Martinez, Chu, Jia, Sakai, Lu,
  Gautam, O’Neil, Becklin, Morris, Matthews, Nishiyama, Campbell, Chappell,
  Chen, Ciurlo, Dehghanfar, Gallego-Cano, Kerzendorf, Lyke, Naoz, Saida,
  Schödel, Takahashi, Takamori, Witzel, \& Wizinowich}]{Do_2019}
Do, T., Hees, A., Ghez, A., {et~al.} 2019, Science, 365, 664,
  \dodoi{10.1126/science.aav8137}

\bibitem[{{Doeleman} {et~al.}(2019){Doeleman}, {Blackburn}, {Dexter}, {Gomez},
  {Johnson}, {Palumbo}, {Weintroub}, {Farah}, {Fish}, {Loinard}, {Lonsdale},
  {Narayanan}, {Patel}, {Pesce}, {Raymond}, {Tilanus}, {Wielgus}, {Akiyama},
  {Bower}, {Broderick}, {Deane}, {Fromm}, {Gammie}, {Gold}, {Janssen},
  {Kawashima}, {Krichbaum}, {Marrone}, {Matthews}, {Mizuno}, {Rezzolla},
  {Roelofs}, {Ros}, {Savolainen}, {Yuan}, {Zhao}, {Blackburn}, {Doeleman},
  {Dexter}, {Gomez}, {Johnson}, {Palumbo}, {Weintroub}, {Farah}, {Fish},
  {Loinard}, {Lonsdale}, {Narayanan}, {Patel}, {Pesce}, {Raymond}, {Tilanus},
  {Wielgus}, {Akiyama}, {Bower}, {Broderick}, {Deane}, {Fromm}, {Gammie},
  {Gold}, {Janssen}, {Kawashima}, {Krichbaum}, {Marrone}, {Matthews}, {Mizuno},
  {Rezzolla}, {Roelofs}, {Ros}, {Savolainen}, {Yuan}, \&
  {Zhao}}]{ngEHT_whitepaper}
{Doeleman}, S., {Blackburn}, L., {Dexter}, J., {et~al.} 2019, \baas, 51, 256.
\newblock \url{https://ui.adsabs.harvard.edu/abs/2019BAAS...51g.256D}

\bibitem[{Emami {et~al.}(2023)Emami, Tiede, Doeleman, Roelofs, Wielgus,
  Blackburn, Liska, Chatterjee, Ripperda, Fuentes, Broderick, Hernquist,
  Alcock, Narayan, Smith, Tremblay, Ricarte, Sun, Anantua, Kovalev, Natarajan,
  \& Vogelsberger}]{Emami_2022}
Emami, R., Tiede, P., Doeleman, S.~S., {et~al.} 2023, Galaxies, 11,
  \dodoi{10.3390/galaxies11010023}

\bibitem[{{Event Horizon Telescope Collaboration}
  {et~al.}(2019{\natexlab{a}}){Event Horizon Telescope Collaboration},
  {Akiyama}, {Alberdi}, {Alef}, {Asada}, {Azulay}, {Baczko}, {Ball},
  {Balokovi{\'c}}, {Barrett}, {Bintley}, {Blackburn}, {Boland}, {Bouman},
  {Bower}, {Bremer}, {Brinkerink}, {Brissenden}, {Britzen}, {Broderick},
  {Broguiere}, {Bronzwaer}, {Byun}, {Carlstrom}, {Chael}, {Chan}, {Chatterjee},
  {Chatterjee}, {Chen}, {Chen}, {Cho}, {Christian}, {Conway}, {Cordes}, {Crew},
  {Cui}, {Davelaar}, {De Laurentis}, {Deane}, {Dempsey}, {Desvignes}, {Dexter},
  {Doeleman}, {Eatough}, {Falcke}, {Fish}, {Fomalont}, {Fraga-Encinas},
  {Freeman}, {Friberg}, {Fromm}, {G{\'o}mez}, {Galison}, {Gammie},
  {Garc{\'\i}a}, {Gentaz}, {Georgiev}, {Goddi}, {Gold}, {Gu}, {Gurwell},
  {Hada}, {Hecht}, {Hesper}, {Ho}, {Ho}, {Honma}, {Huang}, {Huang}, {Hughes},
  {Ikeda}, {Inoue}, {Issaoun}, {James}, {Jannuzi}, {Janssen}, {Jeter}, {Jiang},
  {Johnson}, {Jorstad}, {Jung}, {Karami}, {Karuppusamy}, {Kawashima},
  {Keating}, {Kettenis}, {Kim}, {Kim}, {Kim}, {Kino}, {Koay}, {Koch}, {Koyama},
  {Kramer}, {Kramer}, {Krichbaum}, {Kuo}, {Lauer}, {Lee}, {Li}, {Li},
  {Lindqvist}, {Liu}, {Liuzzo}, {Lo}, {Lobanov}, {Loinard}, {Lonsdale}, {Lu},
  {MacDonald}, {Mao}, {Markoff}, {Marrone}, {Marscher}, {Mart{\'\i}-Vidal},
  {Matsushita}, {Matthews}, {Medeiros}, {Menten}, {Mizuno}, {Mizuno}, {Moran},
  {Moriyama}, {Moscibrodzka}, {M{\"u}ller}, {Nagai}, {Nagar}, {Nakamura},
  {Narayan}, {Narayanan}, {Natarajan}, {Neri}, {Ni}, {Noutsos}, {Okino},
  {Olivares}, {Ortiz-Le{\'o}n}, {Oyama}, {{\"O}zel}, {Palumbo}, {Patel}, {Pen},
  {Pesce}, {Pi{\'e}tu}, {Plambeck}, {PopStefanija}, {Porth}, {Prather},
  {Preciado-L{\'o}pez}, {Psaltis}, {Pu}, {Ramakrishnan}, {Rao}, {Rawlings},
  {Raymond}, {Rezzolla}, {Ripperda}, {Roelofs}, {Rogers}, {Ros}, {Rose},
  {Roshanineshat}, {Rottmann}, {Roy}, {Ruszczyk}, {Ryan}, {Rygl},
  {S{\'a}nchez}, {S{\'a}nchez-Arguelles}, {Sasada}, {Savolainen}, {Schloerb},
  {Schuster}, {Shao}, {Shen}, {Small}, {Sohn}, {SooHoo}, {Tazaki}, {Tiede},
  {Tilanus}, {Titus}, {Toma}, {Torne}, {Trent}, {Trippe}, {Tsuda}, {van
  Bemmel}, {van Langevelde}, {van Rossum}, {Wagner}, {Wardle}, {Weintroub},
  {Wex}, {Wharton}, {Wielgus}, {Wong}, {Wu}, {Young}, {Young}, {Younsi},
  {Yuan}, {Yuan}, {Zensus}, {Zhao}, {Zhao}, {Zhu}, {Algaba}, {Allardi},
  {Amestica}, {Anczarski}, {Bach}, {Baganoff}, {Beaudoin}, {Benson},
  {Berthold}, {Blanchard}, {Blundell}, {Bustamente}, {Cappallo},
  {Castillo-Dom{\'\i}nguez}, {Chang}, {Chang}, {Chang}, {Chen}, {Chilson},
  {Chuter}, {C{\'o}rdova Rosado}, {Coulson}, {Crawford}, {Crowley}, {David},
  {Derome}, {Dexter}, {Dornbusch}, {Dudevoir}, {Dzib}, {Eckart}, {Eckert},
  {Erickson}, {Everett}, {Faber}, {Farah}, {Fath}, {Folkers}, {Forbes},
  {Freund}, {G{\'o}mez-Ruiz}, {Gale}, {Gao}, {Geertsema}, {Graham}, {Greer},
  {Grosslein}, {Gueth}, {Haggard}, {Halverson}, {Han}, {Han}, {Hao},
  {Hasegawa}, {Henning}, {Hern{\'a}ndez-G{\'o}mez}, {Herrero-Illana},
  {Heyminck}, {Hirota}, {Hoge}, {Huang}, {Impellizzeri}, {Jiang}, {Kamble},
  {Keisler}, {Kimura}, {Kono}, {Kubo}, {Kuroda}, {Lacasse}, {Laing}, {Leitch},
  {Li}, {Lin}, {Liu}, {Liu}, {Lu}, {Marson}, {Martin-Cocher}, {Massingill},
  {Matulonis}, {McColl}, {McWhirter}, {Messias}, {Meyer-Zhao}, {Michalik},
  {Monta{\~n}a}, {Montgomerie}, {Mora-Klein}, {Muders}, {Nadolski}, {Navarro},
  {Neilsen}, {Nguyen}, {Nishioka}, {Norton}, {Nowak}, {Nystrom}, {Ogawa},
  {Oshiro}, {Oyama}, {Parsons}, {Paine}, {Pe{\~n}alver}, {Phillips}, {Poirier},
  {Pradel}, {Primiani}, {Raffin}, {Rahlin}, {Reiland}, {Risacher}, {Ruiz},
  {S{\'a}ez-Mada{\'\i}n}, {Sassella}, {Schellart}, {Shaw}, {Silva}, {Shiokawa},
  {Smith}, {Snow}, {Souccar}, {Sousa}, {Sridharan}, {Srinivasan}, {Stahm},
  {Stark}, {Story}, {Timmer}, {Vertatschitsch}, {Walther}, {Wei}, {Whitehorn},
  {Whitney}, {Woody}, {Wouterloot}, {Wright}, {Yamaguchi}, {Yu}, {Zeballos},
  {Zhang}, \& {Ziurys}}]{M87PaperI}
{Event Horizon Telescope Collaboration}, {Akiyama}, K., {Alberdi}, A., {et~al.}
  2019{\natexlab{a}}, \apjl, 875, L1, \dodoi{10.3847/2041-8213/ab0ec7}

\bibitem[{{Event Horizon Telescope Collaboration}
  {et~al.}(2019{\natexlab{b}}){Event Horizon Telescope Collaboration},
  {Akiyama}, {Alberdi}, {Alef}, {Asada}, {Azulay}, {Baczko}, {Ball},
  {Balokovi{\'c}}, {Barrett}, {Bintley}, {Blackburn}, {Boland}, {Bouman},
  {Bower}, {Bremer}, {Brinkerink}, {Brissenden}, {Britzen}, {Broderick},
  {Broguiere}, {Bronzwaer}, {Byun}, {Carlstrom}, {Chael}, {Chan}, {Chatterjee},
  {Chatterjee}, {Chen}, {Chen}, {Cho}, {Christian}, {Conway}, {Cordes}, {Crew},
  {Cui}, {Davelaar}, {De Laurentis}, {Deane}, {Dempsey}, {Desvignes}, {Dexter},
  {Doeleman}, {Eatough}, {Falcke}, {Fish}, {Fomalont}, {Fraga-Encinas},
  {Friberg}, {Fromm}, {G{\'o}mez}, {Galison}, {Gammie}, {Garc{\'\i}a},
  {Gentaz}, {Georgiev}, {Goddi}, {Gold}, {Gu}, {Gurwell}, {Hada}, {Hecht},
  {Hesper}, {Ho}, {Ho}, {Honma}, {Huang}, {Huang}, {Hughes}, {Ikeda}, {Inoue},
  {Issaoun}, {James}, {Jannuzi}, {Janssen}, {Jeter}, {Jiang}, {Johnson},
  {Jorstad}, {Jung}, {Karami}, {Karuppusamy}, {Kawashima}, {Keating},
  {Kettenis}, {Kim}, {Kim}, {Kim}, {Kino}, {Koay}, {Koch}, {Koyama}, {Kramer},
  {Kramer}, {Krichbaum}, {Kuo}, {Lauer}, {Lee}, {Li}, {Li}, {Lindqvist}, {Liu},
  {Liuzzo}, {Lo}, {Lobanov}, {Loinard}, {Lonsdale}, {Lu}, {MacDonald}, {Mao},
  {Markoff}, {Marrone}, {Marscher}, {Mart{\'\i}-Vidal}, {Matsushita},
  {Matthews}, {Medeiros}, {Menten}, {Mizuno}, {Mizuno}, {Moran}, {Moriyama},
  {Moscibrodzka}, {Mul{\ensuremath{\ddot{}}}ler}, {Nagai}, {Nagar}, {Nakamura},
  {Narayan}, {Narayanan}, {Natarajan}, {Neri}, {Ni}, {Noutsos}, {Okino},
  {Olivares}, {Oyama}, {{\"O}zel}, {Palumbo}, {Patel}, {Pen}, {Pesce},
  {Pi{\'e}tu}, {Plambeck}, {PopStefanija}, {Porth}, {Prather},
  {Preciado-L{\'o}pez}, {Psaltis}, {Pu}, {Ramakrishnan}, {Rao}, {Rawlings},
  {Raymond}, {Rezzolla}, {Ripperda}, {Roelofs}, {Rogers}, {Ros}, {Rose},
  {Roshanineshat}, {Rottmann}, {Roy}, {Ruszczyk}, {Ryan}, {Rygl},
  {S{\'a}nchez}, {S{\'a}nchez-Arguelles}, {Sasada}, {Savolainen}, {Schloerb},
  {Schuster}, {Shao}, {Shen}, {Small}, {Sohn}, {SooHoo}, {Tazaki}, {Tiede},
  {Tilanus}, {Titus}, {Toma}, {Torne}, {Trent}, {Trippe}, {Tsuda}, {van
  Bemmel}, {van Langevelde}, {van Rossum}, {Wagner}, {Wardle}, {Weintroub},
  {Wex}, {Wharton}, {Wielgus}, {Wong}, {Wu}, {Young}, {Young}, {Younsi},
  {Yuan}, {Yuan}, {Zensus}, {Zhao}, {Zhao}, {Zhu}, {Anczarski}, {Baganoff},
  {Eckart}, {Farah}, {Haggard}, {Meyer-Zhao}, {Michalik}, {Nadolski},
  {Neilsen}, {Nishioka}, {Nowak}, {Pradel}, {Primiani}, {Souccar},
  {Vertatschitsch}, {Yamaguchi}, \& {Zhang}}]{M87PaperV}
---. 2019{\natexlab{b}}, \apjl, 875, L5, \dodoi{10.3847/2041-8213/ab0f43}

\bibitem[{{Event Horizon Telescope Collaboration} {et~al.}(2021){Event Horizon
  Telescope Collaboration}, {Akiyama}, {Algaba}, {Alberdi}, {Alef}, {Anantua},
  Asada, Azulay, Baczko, Ball, Baloković, Barrett, Benson, Bintley, Blackburn,
  Blundell, Boland, Bouman, Bower, Boyce, Bremer, Brinkerink, Brissenden,
  Britzen, Broderick, Broguiere, Bronzwaer, Byun, Carlstrom, Chael, kwan Chan,
  Chatterjee, Chatterjee, Chen, Chen, Chesler, Cho, Christian, Conway, Cordes,
  Crawford, Crew, Cruz-Osorio, Cui, Davelaar, Laurentis, Deane, Dempsey,
  Desvignes, Dexter, Doeleman, Eatough, Falcke, Farah, Fish, Fomalont, Ford,
  Fraga-Encinas, Friberg, Fromm, Fuentes, Galison, Gammie, García, Gelles,
  Gentaz, Georgiev, Goddi, Gold, Gómez, Gómez-Ruiz, Gu, Gurwell, Hada,
  Haggard, Hecht, Hesper, Himwich, Ho, Ho, Honma, Huang, Huang, Hughes, Ikeda,
  Inoue, Issaoun, James, Jannuzi, Janssen, Jeter, Jiang, Jimenez-Rosales,
  Johnson, Jorstad, Jung, Karami, Karuppusamy, Kawashima, Keating, Kettenis,
  Kim, Kim, Kim, Kim, Kino, Koay, Kofuji, Koch, Koyama, Kramer, Kramer,
  Krichbaum, Kuo, Lauer, Lee, Levis, Li, Li, Lindqvist, Lico, Lindahl, Liu,
  Liu, Liuzzo, Lo, Lobanov, Loinard, Lonsdale, Lu, MacDonald, Mao, Marchili,
  Markoff, Marrone, Marscher, Martí-Vidal, Matsushita, Matthews, Medeiros,
  Menten, Mizuno, Mizuno, Moran, Moriyama, Moscibrodzka, Müller, Musoke,
  Mejías, Michalik, Nadolski, Nagai, Nagar, Nakamura, Narayan, Narayanan,
  Natarajan, Nathanail, Neilsen, Neri, Ni, Noutsos, Nowak, Okino, Olivares,
  Ortiz-León, Oyama, Özel, Palumbo, Park, Patel, Pen, Pesce, Piétu,
  Plambeck, PopStefanija, Porth, Pötzl, Prather, Preciado-López, Psaltis, Pu,
  Ramakrishnan, Rao, Rawlings, Raymond, Rezzolla, Ricarte, Ripperda, Roelofs,
  Rogers, Ros, Rose, Roshanineshat, Rottmann, Roy, Ruszczyk, Rygl, Sánchez,
  Sánchez-Arguelles, Sasada, Savolainen, Schloerb, Schuster, Shao, Shen,
  Small, Sohn, SooHoo, Sun, Tazaki, Tetarenko, Tiede, Tilanus, Titus, Toma,
  Torne, Trent, Traianou, Trippe, van Bemmel, van Langevelde, van Rossum,
  Wagner, Ward-Thompson, Wardle, Weintroub, Wex, Wharton, Wielgus, Wong, Wu,
  Yoon, Young, Young, Younsi, Yuan, Yuan, Zensus, Zhao, Zhao, \&
  Collaboration}]{M87PaperVIII}
{Event Horizon Telescope Collaboration}, {Akiyama}, K., {Algaba}, J.~C.,
  {et~al.} 2021, \apjl, 910, L13, \dodoi{10.3847/2041-8213/abe4de}

\bibitem[{{Event Horizon Telescope Collaboration}
  {et~al.}(2022{\natexlab{a}}){Event Horizon Telescope Collaboration},
  {Akiyama}, {Alberdi}, {Alef}, {Algaba}, {Anantua}, Asada, Azulay, Bach,
  Baczko, Ball, Baloković, Barrett, Bauböck, Benson, Bintley, Blackburn,
  Blundell, Bouman, Bower, Boyce, Bremer, Brinkerink, Brissenden, Britzen,
  Broderick, Broguiere, Bronzwaer, Bustamante, Byun, Carlstrom, Ceccobello,
  Chael, kwan Chan, Chatterjee, Chatterjee, Chen, Chen, Cheng, Cho, Christian,
  Conroy, Conway, Cordes, Crawford, Crew, Cruz-Osorio, Cui, Davelaar,
  Laurentis, Deane, Dempsey, Desvignes, Dexter, Dhruv, Doeleman, Dougal, Dzib,
  Eatough, Emami, Falcke, Farah, Fish, Fomalont, Ford, Fraga-Encinas, Freeman,
  Friberg, Fromm, Fuentes, Galison, Gammie, García, Gentaz, Georgiev, Goddi,
  Gold, Gómez-Ruiz, Gómez, Gu, Gurwell, Hada, Haggard, Haworth, Hecht,
  Hesper, Heumann, Ho, Ho, Honma, Huang, Huang, Hughes, Ikeda, Impellizzeri,
  Inoue, Issaoun, James, Jannuzi, Janssen, Jeter, Jiang, Jiménez-Rosales,
  Johnson, Jorstad, Joshi, Jung, Karami, Karuppusamy, Kawashima, Keating,
  Kettenis, Kim, Kim, Kim, Kim, Kino, Koay, Kocherlakota, Kofuji, Koch, Koyama,
  Kramer, Kramer, Krichbaum, Kuo, Bella, Lauer, Lee, Lee, Leung, Levis, Li,
  Lico, Lindahl, Lindqvist, Lisakov, Liu, Liu, Liuzzo, Lo, Lobanov, Loinard,
  Lonsdale, Lu, Mao, Marchili, Markoff, Marrone, Marscher, Martí-Vidal,
  Matsushita, Matthews, Medeiros, Menten, Michalik, Mizuno, Mizuno, Moran,
  Moriyama, Moscibrodzka, Müller, Mus, Musoke, Myserlis, Nadolski, Nagai,
  Nagar, Nakamura, Narayan, Narayanan, Natarajan, Nathanail, Fuentes, Neilsen,
  Neri, Ni, Noutsos, Nowak, Oh, Okino, Olivares, Ortiz-León, Oyama, Özel,
  Palumbo, Paraschos, Park, Parsons, Patel, Pen, Pesce, Piétu, Plambeck,
  PopStefanija, Porth, Pötzl, Prather, Preciado-López, Psaltis, Pu,
  Ramakrishnan, Rao, Rawlings, Raymond, Rezzolla, Ricarte, Ripperda, Roelofs,
  Rogers, Ros, Romero-Cañizales, Roshanineshat, Rottmann, Roy, Ruiz, Ruszczyk,
  Rygl, Sánchez, Sánchez-Argüelles, Sánchez-Portal, Sasada, Satapathy,
  Savolainen, Schloerb, Schonfeld, Schuster, Shao, Shen, Small, Sohn, SooHoo,
  Souccar, Sun, Tazaki, Tetarenko, Tiede, Tilanus, Titus, Torne, Traianou,
  Trent, Trippe, Turk, van Bemmel, van Langevelde, van Rossum, Vos, Wagner,
  Ward-Thompson, Wardle, Weintroub, Wex, Wharton, Wielgus, Wiik, Witzel,
  Wondrak, Wong, Wu, Yamaguchi, Yoon, Young, Young, Younsi, Yuan, Yuan, Zensus,
  Zhang, Zhao, Zhao, Agurto, Allardi, Amestica, Araneda, Arriagada, Berghuis,
  Bertarini, Berthold, Blanchard, Brown, Cárdenas, Cantzler, Caro,
  Castillo-Domínguez, Chan, Chang, Chang, Chang, Chang, Chen, Chilson, Chuter,
  Ciechanowicz, Colin-Beltran, Coulson, Crowley, Degenaar, Dornbusch, Durán,
  Everett, Faber, Forster, Fuchs, Gale, Geertsema, González, Graham, Gueth,
  Halverson, Han, Han, Hasegawa, Hernández-Rebollar, Herrera, Herrero-Illana,
  Heyminck, Hirota, Hoge, Schimpf, Howie, Huang, Jiang, Jinchi, John, Kimura,
  Klein, Kubo, Kuroda, Kwon, Lacasse, Laing, Leitch, Li, Liu, Liu, Lin, Lu,
  Mac-Auliffe, Martin-Cocher, Matulonis, Maute, Messias, Meyer-Zhao, Montaña,
  Montenegro-Montes, Montgomerie, Nolasco, Muders, Nishioka, Norton, Nystrom,
  Ogawa, Olivares, Oshiro, Pérez-Beaupuits, Parra, Phillips, Poirier, Pradel,
  Qiu, Raffin, Rahlin, Ramírez, Ressler, Reynolds, Rodríguez-Montoya,
  Saez-Madain, Santana, Shaw, Shirkey, Silva, Snow, Sousa, Sridharan, Stahm,
  Stark, Test, Torstensson, Venegas, Walther, Wei, White, Wieching, Wijnands,
  Wouterloot, Yu, (于威), \& Zeballos}]{SgrAPaperI}
{Event Horizon Telescope Collaboration}, {Akiyama}, K., {Alberdi}, A., {et~al.}
  2022{\natexlab{a}}, \apjl, 930, L12, \dodoi{10.3847/2041-8213/ac6674}

\bibitem[{{Event Horizon Telescope Collaboration}
  {et~al.}(2022{\natexlab{b}}){Event Horizon Telescope Collaboration},
  {Akiyama}, {Alberdi}, {Alef}, {Algaba}, {Anantua}, Asada, Azulay, Bach,
  Baczko, Ball, Baloković, Barrett, Bauböck, Benson, Bintley, Blackburn,
  Blundell, Bouman, Bower, Boyce, Bremer, Brinkerink, Brissenden, Britzen,
  Broderick, Broguiere, Bronzwaer, Bustamante, Byun, Carlstrom, Ceccobello,
  Chael, kwan Chan, Chatterjee, Chatterjee, Chen, Chen, Cheng, Cho, Christian,
  Conroy, Conway, Cordes, Crawford, Crew, Cruz-Osorio, Cui, Davelaar,
  Laurentis, Deane, Dempsey, Desvignes, Dexter, Dhruv, Doeleman, Dougal, Dzib,
  Eatough, Emami, Falcke, Farah, Fish, Fomalont, Ford, Fraga-Encinas, Freeman,
  Friberg, Fromm, Fuentes, Galison, Gammie, García, Gentaz, Georgiev, Goddi,
  Gold, Gómez-Ruiz, Gómez, Gu, Gurwell, Hada, Haggard, Haworth, Hecht,
  Hesper, Heumann, Ho, Ho, Honma, Huang, Huang, Hughes, Ikeda, Impellizzeri,
  Inoue, Issaoun, James, Jannuzi, Janssen, Jeter, Jiang, Jiménez-Rosales,
  Johnson, Jorstad, Joshi, Jung, Karami, Karuppusamy, Kawashima, Keating,
  Kettenis, Kim, Kim, Kim, Kim, Kino, Koay, Kocherlakota, Kofuji, Koch, Koyama,
  Kramer, Kramer, Krichbaum, Kuo, Bella, Lauer, Lee, Lee, Leung, Levis, Li,
  Lico, Lindahl, Lindqvist, Lisakov, Liu, Liu, Liuzzo, Lo, Lobanov, Loinard,
  Lonsdale, Lu, Mao, Marchili, Markoff, Marrone, Marscher, Martí-Vidal,
  Matsushita, Matthews, Medeiros, Menten, Michalik, Mizuno, Mizuno, Moran,
  Moriyama, Moscibrodzka, Müller, Mus, Musoke, Myserlis, Nadolski, Nagai,
  Nagar, Nakamura, Narayan, Narayanan, Natarajan, Nathanail, Fuentes, Neilsen,
  Neri, Ni, Noutsos, Nowak, Oh, Okino, Olivares, Ortiz-León, Oyama, Özel,
  Palumbo, Paraschos, Park, Parsons, Patel, Pen, Pesce, Piétu, Plambeck,
  PopStefanija, Porth, Pötzl, Prather, Preciado-López, Psaltis, Pu,
  Ramakrishnan, Rao, Rawlings, Raymond, Rezzolla, Ricarte, Ripperda, Roelofs,
  Rogers, Ros, Romero-Cañizales, Roshanineshat, Rottmann, Roy, Ruiz, Ruszczyk,
  Rygl, Sánchez, Sánchez-Argüelles, Sánchez-Portal, Sasada, Satapathy,
  Savolainen, Schloerb, Schonfeld, Schuster, Shao, Shen, Small, Sohn, SooHoo,
  Souccar, Sun, Tazaki, Tetarenko, Tiede, Tilanus, Titus, Torne, Traianou,
  Trent, Trippe, Turk, van Bemmel, van Langevelde, van Rossum, Vos, Wagner,
  Ward-Thompson, Wardle, Weintroub, Wex, Wharton, Wielgus, Wiik, Witzel,
  Wondrak, Wong, Wu, Yamaguchi, Yoon, Young, Young, Younsi, Yuan, Yuan, Zensus,
  Zhang, Zhao, Zhao, Chan, Qiu, Ressler, \& White}]{SgrAPaperV}
---. 2022{\natexlab{b}}, \apjl, 930, L16, \dodoi{10.3847/2041-8213/ac6672}

\bibitem[{{Event Horizon Telescope Collaboration}
  {et~al.}(2022{\natexlab{c}}){Event Horizon Telescope Collaboration},
  {Akiyama}, {Alberdi}, {Alef}, {Algaba}, {Anantua}, Asada, Azulay, Bach,
  Baczko, Ball, Baloković, Barrett, Bauböck, Benson, Bintley, Blackburn,
  Blundell, Bouman, Bower, Boyce, Bremer, Brinkerink, Brissenden, Britzen,
  Broderick, Broguiere, Bronzwaer, Bustamante, Byun, Carlstrom, Ceccobello,
  Chael, kwan Chan, Chatterjee, Chatterjee, Chen, Chen, Cheng, Cho, Christian,
  Conroy, Conway, Cordes, Crawford, Crew, Cruz-Osorio, Cui, Davelaar,
  Laurentis, Deane, Dempsey, Desvignes, Dexter, Dhruv, Doeleman, Dougal, Dzib,
  Eatough, Emami, Falcke, Farah, Fish, Fomalont, Ford, Fraga-Encinas, Freeman,
  Friberg, Fromm, Fuentes, Galison, Gammie, García, Gentaz, Georgiev, Goddi,
  Gold, Gómez-Ruiz, Gómez, Gu, Gurwell, Hada, Haggard, Haworth, Hecht,
  Hesper, Heumann, Ho, Ho, Honma, Huang, Huang, Hughes, Ikeda, Impellizzeri,
  Inoue, Issaoun, James, Jannuzi, Janssen, Jeter, Jiang, Jiménez-Rosales,
  Johnson, Jorstad, Joshi, Jung, Karami, Karuppusamy, Kawashima, Keating,
  Kettenis, Kim, Kim, Kim, Kim, Kino, Koay, Kocherlakota, Kofuji, Koch, Koyama,
  Kramer, Kramer, Krichbaum, Kuo, Bella, Lauer, Lee, Lee, Leung, Levis, Li,
  Lico, Lindahl, Lindqvist, Lisakov, Liu, Liu, Liuzzo, Lo, Lobanov, Loinard,
  Lonsdale, Lu, Mao, Marchili, Markoff, Marrone, Marscher, Martí-Vidal,
  Matsushita, Matthews, Medeiros, Menten, Michalik, Mizuno, Mizuno, Moran,
  Moriyama, Moscibrodzka, Müller, Mus, Musoke, Myserlis, Nadolski, Nagai,
  Nagar, Nakamura, Narayan, Narayanan, Natarajan, Nathanail, Fuentes, Neilsen,
  Neri, Ni, Noutsos, Nowak, Oh, Okino, Olivares, Ortiz-León, Oyama, Palumbo,
  Paraschos, Park, Parsons, Patel, Pen, Pesce, Piétu, Plambeck, PopStefanija,
  Porth, Pötzl, Prather, Preciado-López, Pu, Ramakrishnan, Rao, Rawlings,
  Raymond, Rezzolla, Ricarte, Ripperda, Roelofs, Rogers, Ros,
  Romero-Cañizales, Roshanineshat, Rottmann, Roy, Ruiz, Ruszczyk, Rygl,
  Sánchez, Sánchez-Argüelles, Sánchez-Portal, Sasada, Satapathy,
  Savolainen, Schloerb, Schonfeld, Schuster, Shao, Shen, Small, Sohn, SooHoo,
  Souccar, Sun, Tazaki, Tetarenko, Tiede, Tilanus, Titus, Torne, Traianou,
  Trent, Trippe, Turk, van Bemmel, van Langevelde, van Rossum, Vos, Wagner,
  Ward-Thompson, Wardle, Weintroub, Wex, Wharton, Wielgus, Wiik, Witzel,
  Wondrak, Wong, Wu, Yamaguchi, Yoon, Young, Young, Younsi, Yuan, Yuan, Zensus,
  Zhang, Zhao, Zhao, \& Chang}]{SgrAPaperIV}
---. 2022{\natexlab{c}}, \apjl, 930, L15, \dodoi{10.3847/2041-8213/ac6736}

\bibitem[{{Gammie} {et~al.}(2003){Gammie}, {McKinney}, \&
  {T{\'o}th}}]{Gammie_03}
{Gammie}, C.~F., {McKinney}, J.~C., \& {T{\'o}th}, G. 2003, \apj, 589, 444,
  \dodoi{10.1086/374594}

\bibitem[{Gebhardt {et~al.}(2011)Gebhardt, Adams, Richstone, Lauer, Faber,
  Gültekin, Murphy, \& Tremaine}]{Gebhardt_2011}
Gebhardt, K., Adams, J., Richstone, D., {et~al.} 2011, \apj, 729, 119,
  \dodoi{10.1088/0004-637X/729/2/119}

\bibitem[{Ghez {et~al.}(2003)Ghez, Duchêne, Matthews, Hornstein, Tanner,
  Larkin, Morris, Becklin, Salim, Kremenek, Thompson, Soifer, Neugebauer, \&
  McLean}]{Ghez_2003}
Ghez, A.~M., Duchêne, G., Matthews, K., {et~al.} 2003, \apjl, 586, L127,
  \dodoi{10.1086/374804}

\bibitem[{Ghez {et~al.}(2008)Ghez, Salim, Weinberg, Lu, Do, Dunn, Matthews,
  Morris, Yelda, Becklin, Kremenek, Milosavljevic, \& Naiman}]{Ghez_2008}
Ghez, A.~M., Salim, S., Weinberg, N.~N., {et~al.} 2008, \apjl, 689, 1044,
  \dodoi{10.1086/592738}

\bibitem[{Gralla \& Lupsasca(2020)}]{Gralla_2020}
Gralla, S.~E., \& Lupsasca, A. 2020, PhRvD, 102, 124003,
  \dodoi{10.1103/PhysRevD.102.124003}

\bibitem[{{GRAVITY Collaboration} {et~al.}(2018){GRAVITY Collaboration},
  Abuter, Amorim, Bauböck, Berger, Bonnet, Brandner, Cl{\'{e} }net,
  du~Foresto, de~Zeeuw, Deen, Dexter, Duvert, Eckart, Eisenhauer, Schreiber,
  Garcia, Gao, Gendron, Genzel, Gillessen, Guajardo, Habibi, Haubois, Henning,
  Hippler, Horrobin, Huber, Jim{\'{e}}nez-Rosales, Jocou, Kervella, Lacour,
  Lapeyr{\`{e}}re, Lazareff, Bouquin, L{\'{e}}na, Lippa, Ott, Panduro, Paumard,
  Perraut, Perrin, Pfuhl, Plewa, Rabien, Rodr{\'{\i}}guez-Coira, Rousset,
  Sternberg, Straub, Straubmeier, Sturm, Tacconi, Vincent, von Fellenberg,
  Waisberg, Widmann, Wieprecht, Wiezorrek, Woillez, \& Yazici}]{GRAVITY_2018}
{GRAVITY Collaboration}, Abuter, R., Amorim, A., {et~al.} 2018, \aap, 618, L10,
  \dodoi{10.1051/0004-6361/201834294}

\bibitem[{{GRAVITY Collaboration} {et~al.}(2019){GRAVITY Collaboration},
  {Abuter, R.}, {Amorim, A.}, {Baub\"ock, M.}, {Berger, J. P.}, {Bonnet, H.},
  {Brandner, W.}, {Cl\'enet, Y.}, {Coud\'e du Foresto, V.}, {de Zeeuw, P. T.},
  {Dexter, J.}, {Duvert, G.}, {Eckart, A.}, {Eisenhauer, F.}, {F\"orster
  Schreiber, N. M.}, {Garcia, P.}, {Gao, F.}, {Gendron, E.}, {Genzel, R.},
  {Gerhard, O.}, {Gillessen, S.}, {Habibi, M.}, {Haubois, X.}, {Henning, T.},
  {Hippler, S.}, {Horrobin, M.}, {Jim\'enez-Rosales, A.}, {Jocou, L.},
  {Kervella, P.}, {Lacour, S.}, {Lapeyr\`ere, V.}, {Le Bouquin, J.-B.},
  {L\'ena, P.}, {Ott, T.}, {Paumard, T.}, {Perraut, K.}, {Perrin, G.}, {Pfuhl,
  O.}, {Rabien, S.}, {Rodriguez Coira, G.}, {Rousset, G.}, {Scheithauer, S.},
  {Sternberg, A.}, {Straub, O.}, {Straubmeier, C.}, {Sturm, E.}, {Tacconi, L.
  J.}, {Vincent, F.}, {von Fellenberg, S.}, {Waisberg, I.}, {Widmann, F.},
  {Wieprecht, E.}, {Wiezorrek, E.}, {Woillez, J.}, \& {Yazici,
  S.}}]{GRAVITY_2019}
{GRAVITY Collaboration}, {Abuter, R.}, {Amorim, A.}, {et~al.} 2019, A\&A, 625,
  L10, \dodoi{10.1051/0004-6361/201935656}

\bibitem[{{GRAVITY Collaboration} {et~al.}(2020{\natexlab{a}}){GRAVITY
  Collaboration}, {Abuter}, {Amorim}, {Baub{\"o}ck}, {Berger}, {Bonnet},
  {Brandner}, {Cardoso}, {Cl{\'e}net}, {de Zeeuw}, {Dexter}, {Eckart},
  {Eisenhauer}, {F{\"o}rster Schreiber}, {Garcia}, {Gao}, {Gendron}, {Genzel},
  {Gillessen}, {Habibi}, {Haubois}, {Henning}, {Hippler}, {Horrobin},
  {Jim{\'e}nez-Rosales}, {Jochum}, {Jocou}, {Kaufer}, {Kervella}, {Lacour},
  {Lapeyr{\`e}re}, {Le Bouquin}, {L{\'e}na}, {Nowak}, {Ott}, {Paumard},
  {Perraut}, {Perrin}, {Pfuhl}, {Rodr{\'\i}guez-Coira}, {Shangguan},
  {Scheithauer}, {Stadler}, {Straub}, {Straubmeier}, {Sturm}, {Tacconi},
  {Vincent}, {von Fellenberg}, {Waisberg}, {Widmann}, {Wieprecht}, {Wiezorrek},
  {Woillez}, {Yazici}, \& {Zins}}]{GRAVITY_2020_precession}
{GRAVITY Collaboration}, {Abuter}, R., {Amorim}, A., {et~al.}
  2020{\natexlab{a}}, \aap, 636, L5, \dodoi{10.1051/0004-6361/202037813}

\bibitem[{{GRAVITY Collaboration} {et~al.}(2020{\natexlab{b}}){GRAVITY
  Collaboration}, {Baub{\"o}ck}, {Dexter}, {Abuter}, {Amorim}, {Berger},
  {Bonnet}, {Brandner}, {Cl{\'e}net}, {Coud{\'e} Du Foresto}, {de Zeeuw},
  {Duvert}, {Eckart}, {Eisenhauer}, {F{\"o}rster Schreiber}, {Gao}, {Garcia},
  {Gendron}, {Genzel}, {Gerhard}, {Gillessen}, {Habibi}, {Haubois}, {Henning},
  {Hippler}, {Horrobin}, {Jim{\'e}nez-Rosales}, {Jocou}, {Kervella}, {Lacour},
  {Lapeyr{\`e}re}, {Le Bouquin}, {L{\'e}na}, {Ott}, {Paumard}, {Perraut},
  {Perrin}, {Pfuhl}, {Rabien}, {Rodriguez Coira}, {Rousset}, {Scheithauer},
  {Stadler}, {Sternberg}, {Straub}, {Straubmeier}, {Sturm}, {Tacconi},
  {Vincent}, {von Fellenberg}, {Waisberg}, {Widmann}, {Wieprecht}, {Wiezorrek},
  {Woillez}, \& {Yazici}}]{GRAVITY_2020_flares}
{GRAVITY Collaboration}, {Baub{\"o}ck}, M., {Dexter}, J., {et~al.}
  2020{\natexlab{b}}, \aap, 635, A143, \dodoi{10.1051/0004-6361/201937233}

\bibitem[{Johnson {et~al.}(2019)Johnson, Haworth, Pesce, Palumbo, Blackburn,
  Akiyama, Boroson, Bouman, Farah, Fish, Honma, Kawashima, Kino, Raymond,
  Silver, Weintroub, Wielgus, Doeleman, Kauffmann, Keating, Krichbaum, Loinard,
  Narayanan, Doi, James, Marrone, Mizuno, \& Nagai}]{ngEHT_whitepaper2}
Johnson, M., Haworth, K., Pesce, D.~W., {et~al.} 2019, \baas, 51, 235.
\newblock \url{https://ui.adsabs.harvard.edu/abs/2019BAAS...51g.256D/abstract}

\bibitem[{{Mo{\'s}cibrodzka} \& {Gammie}(2018)}]{Mosc_2018}
{Mo{\'s}cibrodzka}, M., \& {Gammie}, C.~F. 2018, \mnras, 475, 43,
  \dodoi{10.1093/mnras/stx3162}

\bibitem[{Ressler {et~al.}(2020)Ressler, White, Quataert, \&
  Stone}]{Ressler_2020}
Ressler, S.~M., White, C.~J., Quataert, E., \& Stone, J.~M. 2020, \apjl, 896,
  L6, \dodoi{10.3847/2041-8213/ab9532}

\bibitem[{{Ricarte} {et~al.}(2022){Ricarte}, {Palumbo}, {Narayan}, {Roelofs},
  \& {Emami}}]{Ricarte_2022}
{Ricarte}, A., {Palumbo}, D. C.~M., {Narayan}, R., {Roelofs}, F., \& {Emami},
  R. 2022, \apjl, 941, L12, \dodoi{10.3847/2041-8213/aca087}

\bibitem[{{Sch{\"o}del} {et~al.}(2002){Sch{\"o}del}, {Ott}, {Genzel},
  {Hofmann}, {Lehnert}, {Eckart}, {Mouawad}, {Alexander}, {Reid}, {Lenzen},
  {Hartung}, {Lacombe}, {Rouan}, {Gendron}, {Rousset}, {Lagrange}, {Brandner},
  {Ageorges}, {Lidman}, {Moorwood}, {Spyromilio}, {Hubin}, \&
  {Menten}}]{Schoedel_2002}
{Sch{\"o}del}, R., {Ott}, T., {Genzel}, R., {et~al.} 2002, \nat, 419, 694,
  \dodoi{10.1038/nature01121}

\bibitem[{{Spruit}(1987)}]{Spruit_1987}
{Spruit}, H.~C. 1987, \aap, 184, 173

\bibitem[{{Vos} {et~al.}(2022){Vos}, {Mo{\'s}cibrodzka}, \&
  {Wielgus}}]{Vos_2022}
{Vos}, J., {Mo{\'s}cibrodzka}, M.~A., \& {Wielgus}, M. 2022, \aap, 668, A185,
  \dodoi{10.1051/0004-6361/202244840}

\bibitem[{White {et~al.}(2020)White, Dexter, Blaes, \& Quataert}]{White_2020}
White, C.~J., Dexter, J., Blaes, O., \& Quataert, E. 2020, \apj, 894, 14,
  \dodoi{10.3847/1538-4357/ab8463}

\bibitem[{Wielgus {et~al.}(2020)Wielgus, Akiyama, Blackburn, kwan Chan, Dexter,
  Doeleman, Fish, Issaoun, Johnson, Krichbaum, Lu, Pesce, Wong, Bower,
  Broderick, Chael, Chatterjee, Gammie, Georgiev, Hada, Loinard, Markoff,
  Marrone, Plambeck, Weintroub, Dexter, MacMahon, Wright, Alberdi, Alef, Asada,
  Azulay, Baczko, Ball, Baloković, Barausse, Barrett, Bintley, Boland, Bouman,
  Bremer, Brinkerink, Brissenden, Britzen, Broguiere, Bronzwaer, Byun,
  Carlstrom, Chatterjee, Chen, Chen, Cho, Christian, Conway, Cordes, Crew, Cui,
  Davelaar, Laurentis, Deane, Dempsey, Desvignes, Dzib, Eatough, Falcke,
  Fomalont, Fraga-Encinas, Friberg, Fromm, Galison, García, Gentaz, Goddi,
  Gold, Gómez, Gómez-Ruiz, Gu, Gurwell, Hecht, Hesper, Ho, Ho, Honma, Huang,
  Huang, Hughes, Inoue, James, Jannuzi, Janssen, Jeter, Jiang, Jimenez-Rosales,
  Jorstad, Jung, Karami, Karuppusamy, Kawashima, Keating, Kettenis, Kim, Kim,
  Kim, Kino, Koay, Koch, Koyama, Kramer, Kramer, Kuo, Lauer, Lee, Li, Li,
  Lindqvist, Lico, Liu, Liuzzo, Lo, Lobanov, Lonsdale, MacDonald, Mao,
  Marchili, Marscher, Martí-Vidal, Matsushita, Matthews, Medeiros, Menten,
  Mizuno, Mizuno, Moran, Moriyama, Moscibrodzka, Müller, Musoke, Nagai, Nagar,
  Nakamura, Narayan, Narayanan, Natarajan, Nathanail, Neri, Ni, Noutsos, Okino,
  Olivares, Ortiz-León, Oyama, Özel, Palumbo, Park, Patel, Pen, Piétu,
  PopStefanija, Porth, Prather, Preciado-López, Psaltis, Pu, Ramakrishnan,
  Rao, Rawlings, Raymond, Rezzolla, Ripperda, Roelofs, Rogers, Ros, Rose,
  Roshanineshat, Rottmann, Roy, Ruszczyk, Ryan, Rygl, Sánchez,
  Sánchez-Arguelles, Sasada, Savolainen, Schloerb, Schuster, Shao, Shen,
  Small, Sohn, SooHoo, Tazaki, Tiede, Tilanus, Titus, Toma, Torne, Trent,
  Traianou, Trippe, Tsuda, van Bemmel, van Langevelde, van Rossum, Wagner,
  Wardle, Ward-Thompson, Wex, Wharton, Wu, Yoon, Young, Young, Younsi, Yuan,
  Yuan, Zensus, Zhao, Zhao, \& Zhu}]{Wielgus_2020}
Wielgus, M., Akiyama, K., Blackburn, L., {et~al.} 2020, \apj, 901, 67,
  \dodoi{10.3847/1538-4357/abac0d}

\bibitem[{Wielgus {et~al.}(2022)Wielgus, Marchili, Martí-Vidal, Keating,
  Ramakrishnan, Tiede, Fomalont, Issaoun, Neilsen, Nowak, Blackburn, Gammie,
  Goddi, Haggard, Lee, Moscibrodzka, Tetarenko, Bower, kwan Chan, Chatterjee,
  Chesler, Dexter, Doeleman, Georgiev, Gurwell, Johnson, Marrone, Mus, Psaltis,
  Ripperda, Witzel, Akiyama, Alberdi, Alef, Algaba, Anantua, Asada, Azulay,
  Bach, Baczko, Ball, Baloković, Barrett, Bauböck, Benson, Bintley, Blundell,
  Boland, Bouman, Boyce, Bremer, Brinkerink, Brissenden, Britzen, Broderick,
  Broguiere, Bronzwaer, Bustamante, Byun, Carlstrom, Ceccobello, Chael,
  Chatterjee, Chen, Chen, Cho, Christian, Conroy, Conway, Cordes, Crawford,
  Crew, Cruz-Osorio, Cui, Davelaar, Laurentis, Deane, Dempsey, Desvignes,
  Dhruv, Dzib, Eatough, Emami, Falcke, Farah, Fish, Ford, Fraga-Encinas,
  Freeman, Friberg, Fromm, Fuentes, Galison, García, Gentaz, Gold,
  Gómez-Ruiz, Gómez, Gu, Hada, Haworth, Hecht, Hesper, Ho, Ho, Honma, Huang,
  Huang, Hughes, Ikeda, Impellizzeri, Inoue, James, Jannuzi, Janssen, Jeter,
  Jiang, Jiménez-Rosales, Jorstad, Joshi, Jung, Karami, Karuppusamy,
  Kawashima, Kettenis, Kim, Kim, Kim, Kim, Kino, Koay, Kocherlakota, Kofuji,
  Koch, Koyama, Kramer, Kramer, Krichbaum, Kuo, Bella, Lauer, Lee, Leung,
  Levis, Li, Lico, Lindahl, Lindqvist, Lisakov, Liu, Liu, Liuzzo, Lo, Lobanov,
  Loinard, Lonsdale, Lu, Mao, Markoff, Marscher, Matsushita, Matthews,
  Medeiros, Menten, Michalik, Mizuno, Mizuno, Moran, Moriyama, Müller, Musoke,
  Myserlis, Nadolski, Nagai, Nagar, Nakamura, Narayan, Narayanan, Natarajan,
  Nathanail, Fuentes, Neri, Ni, Noutsos, Oh, Okino, Olivares, Ortiz-León,
  Oyama, Özel, Palumbo, Paraschos, Park, Parsons, Patel, Pen, Pesce, Piétu,
  Plambeck, PopStefanija, Porth, Pötzl, Prather, Preciado-López, Pu, Rao,
  Rawlings, Raymond, Rezzolla, Ricarte, Roelofs, Rogers, Ros, Romero-Canizales,
  Roshanineshat, Rottmann, Roy, Ruiz, Ruszczyk, Rygl, Sánchez,
  Sánchez-Argüelles, Sánchez-Portal, Sasada, Satapathy, Savolainen,
  Schloerb, Schuster, Shao, Shen, Small, Sohn, SooHoo, Souccar, Sun, Tazaki,
  Tilanus, Titus, Torne, Traianou, Trent, Trippe, van Bemmel, van Langevelde,
  van Rossum, Vos, Wagner, Ward-Thompson, Wardle, Weintroub, Wex, Wharton,
  Wiik, Wondrak, Wong, Wu, Yamaguchi, Yoon, Young, Young, Younsi, Yuan, Yuan,
  Zensus, Zhang, Zhao, \& Zhao}]{Wielgus_2022}
Wielgus, M., Marchili, N., Martí-Vidal, I., {et~al.} 2022, \apjl, 930, L19,
  \dodoi{10.3847/2041-8213/ac6428}

\bibitem[{{Wielgus} {et~al.}(2022){Wielgus}, {Moscibrodzka}, {Vos}, {Gelles},
  {Mart{\'\i}-Vidal}, {Farah}, {Marchili}, {Goddi}, \&
  {Messias}}]{Wielgus_2022_hotspots}
{Wielgus}, M., {Moscibrodzka}, M., {Vos}, J., {et~al.} 2022, \aap, 665, L6,
  \dodoi{10.1051/0004-6361/202244493}

\bibitem[{Wong {et~al.}(2022)Wong, Prather, Dhruv, Ryan, Mościbrodzka, kwan
  Chan, Joshi, Yarza, Ricarte, Shiokawa, Dolence, Noble, McKinney, \&
  Gammie}]{Wong_2022}
Wong, G.~N., Prather, B.~S., Dhruv, V., {et~al.} 2022, \apjs, 259, 64,
  \dodoi{10.3847/1538-4365/ac582e}

\end{thebibliography}
\bibliographystyle{aasjournal}

\end{document}